\newcommand{\planck}{{\em Planck }}
\newcommand{\plancks}{{\em Planck}'s }
\newcommand{\healpix}{{\tt HEALPix }}
\newcommand{\DE}{dark energy }
\newcommand{\dd}{\mathrm{d}}
\newcommand{\cc}{\mathrm{c}}
\newcommand{\e}{\mathrm{e}}
\newcommand{\p}{\mathrm{p}}
\newcommand{\kB}{k_{\mathrm{B}}}
\newcommand{\sT}{\sigma_{\mathrm{T}}}

\documentclass[]{aa}  
\usepackage{graphicx}
\usepackage{subfigure}
\usepackage{txfonts}
\usepackage{natbib}

\bibpunct{(}{)}{;}{a}{}{,}	
\graphicspath {{./figures/}}

\begin{document}

   \title{Impact of early dark energy on the \planck SZ cluster sample}

   \author{Jean-Claude Waizmann \and Matthias Bartelmann}

   \institute{Zentrum f\"ur Astronomie, ITA, Universit\"at Heidelberg,
              Albert-\"Uberle-Strasse 2, 69120 Heidelberg, Germany\\
              \email{waizmann@ita.uni-heidelberg.de, mbartelmann@ita.uni-heidelberg.de}
   }

  \abstract
   {One science goal of the upcoming \planck mission is to perform a full-sky cluster survey based on the Sunyaev-Zel'dovich (SZ) effect, which leads to the question of how such a survey would be affected by cosmological models with a different history of structure formation than \(\Lambda\)CDM. One class of these models are early dark energy (EDE) cosmologies, where the dark energy contribution does not vanish at early times.}
   {Since structures grow slower in the presence of EDE, one expects an increase in the number of galaxy clusters compared to \(\Lambda\)CDM at intermediate and high redshifts, which could explain the reported excess of the angular CMB power spectrum on cluster scales via an enhanced SZ contribution. We study the impact of EDE on \plancks expected cluster sample. }
   {To obtain realistic simulations, we constructed full-sky SZ maps for EDE and \(\Lambda\)CDM cosmologies, taking angular cluster correlation into account. Using these maps, we simulated \planck observations with and without Galactic foregrounds and fed the results into our filter pipeline based on the spherical multi-frequency matched filters.}
   {For the case of EDE cosmologies, we clearly find an increase in the detected number of clusters compared to the fiducial \(\Lambda\)CDM case. This shows that the spherical multi-frequency matched filter is sensitive enough to find deviations from the \(\Lambda\)CDM sample, being caused by EDE. In addition we find an interesting effect of EDE on the completeness of the cluster sample, such that EDE helps to obtain cleaner samples.}
   {}

   \keywords{Galaxies: clusters: general - cosmic microwave background - Methods: numerical - Space vehicles:   \planck}

   \authorrunning{J.-C. Waizmann \& M. Bartelmann} 

   \maketitle

\section{Introduction}

Recent observations with ground based CMB experiments like BIMA, CBI, and ACBAR \citep{2002ApJ...581...86D,2004ApJ...609..498R,2008arXiv0801.1491R} report an excess in angular power at high multipoles \(\ell>2000\) with respect to the theoretically expected CMB power spectrum. If explained by the contribution of the thermal Sunyaev-Zel'dovich (SZ) effect \citep{1972CoASP...4..173S} within the framework of the \(\Lambda\)CDM cosmology, a rather high normalisation of the mass fluctuations \(\sigma_8\sim 1\) would be required \citep{2005ApJ...626...12B}. Such a high value would contradict the \(\sigma_8=0.796\pm{0.036}\) reported by the WMAP-5 results \citep{2008arXiv0803.0586D}. Furthermore, despite being controversial, the observed abundance of gravitational arcs \citep{1998A&A...330....1B,2000MNRAS.314..338M,2003MNRAS.346...67M,2004ApJ...606L..93W,2005ApJ...622...99D,2007ApJ...654..714H,2008arXiv0803.0656F} and their presence in high-redshift galaxy clusters with \(z\gtrsim 1\) \citep{1998A&A...340L..27H,2003ApJ...593...48G,2001A&A...377..778T,2003ApJ...584..691Z}, gives rise to tension with the concordance model.\\
One explanation of these contradictory observations is offered by models of early dark energy (EDE), in which the non-vanishing contribution of a dark-energy density at high redshifts alters structure formation. The effect of a few percent of EDE on non-linear structure formation has been analytically studied for instance by \cite{2006A&A...454...27B}, who find that such a contribution may lead to ten times more galaxy clusters at redshift unity than \(\Lambda\)CDM. In this way the strong-lensing measurements \citep{2007A&A...461...49F, 2008arXiv0803.0656F} and the observed high-multipole CMB power spectrum \citep{2007MNRAS.380..637S} could be explained.

In addition, EDE models, belonging to the class of dynamical dark energy models, also offer an explanation for the present observed cosmic acceleration. Based on the concept of scalar fields \citep{1988PhRvD..37.3406R, 1988NuPhB.302..668W, 1998PhRvL..80.1582C, 1999PhRvD..59b3509L, 1999PhLB..459..570Z} exerting negative pressure, these models are able to reproduce the late-time accelerated expansion of the Universe. The dynamical evolution of such a field, if present at all, is largely unconstrained by present observations of SN Ia, LSS and CMB and leaves room for a range of different classes of models \citep{2005JCAP...11..007D, 2007PhRvD..75b3003D}. Models giving rise to attractor solutions \citep{1988PhRvD..37.3406R, 1988NuPhB.302..668W}, where the evolution of the \DE component follows the dominant component of the cosmic fluid, naturally suggest that \DE might be non-negligible during long periods of the evolution of the Universe.

Because a few percent of EDE significantly change the abundance of galaxy clusters with respect to the \(\Lambda\)CDM case at higher redshifts, we have reason to believe that such an excess could be detectable by the \planck mission. To investigate this we decided to simulate full-sky \planck observations of the SZ effect for models of EDE in order to quantify the impact on the expected cluster sample.\\

Section 2 briefly reviews the concepts of EDE and is followed by a recapitulation of the SZ effect in Sect. 3. The construction of semi-analytic full-sky SZ maps for different cosmologies is explained in Sect. 4. Section 5 outlines how realistic simulations of \planck observations are performed. The method of cluster extraction is discussed in Sect. 6, followed by an exploitation of the obtained cluster sample for the considered cosmological models in Sect. 7 and finished by the conclusions in Sect. 8.
\begin{figure*}
   \centering
	\subfigure{
        	\includegraphics [height=5.6cm] {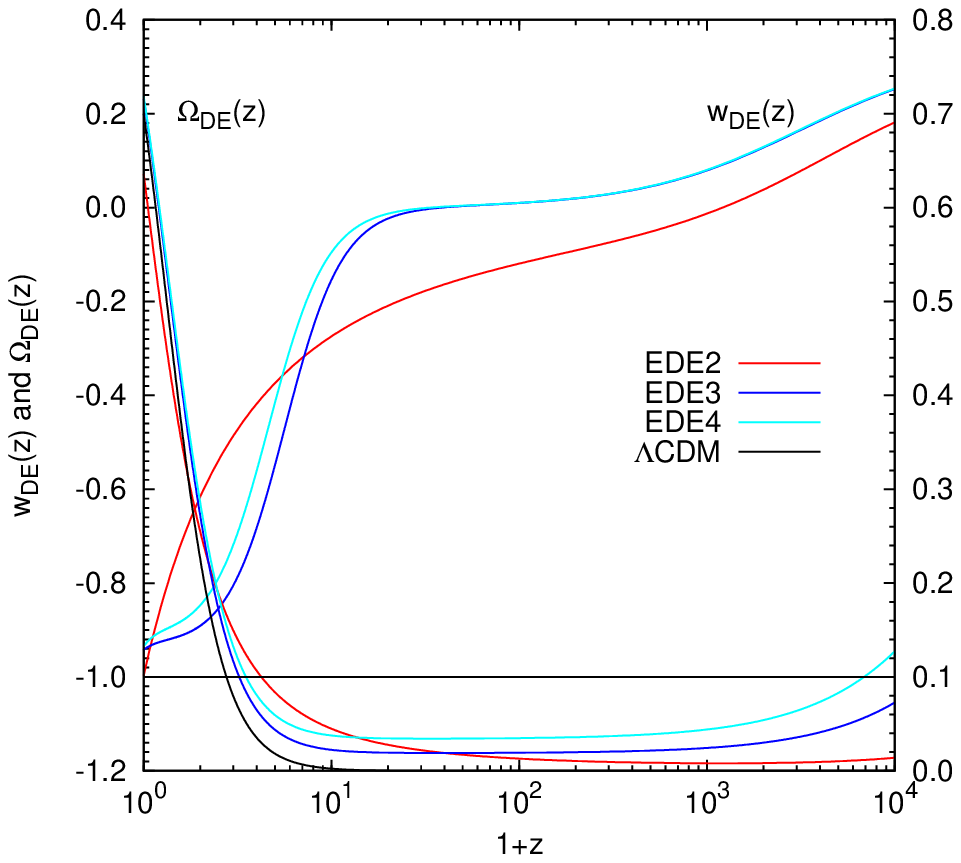}}
	\subfigure{
        	\includegraphics [height=5.6cm] {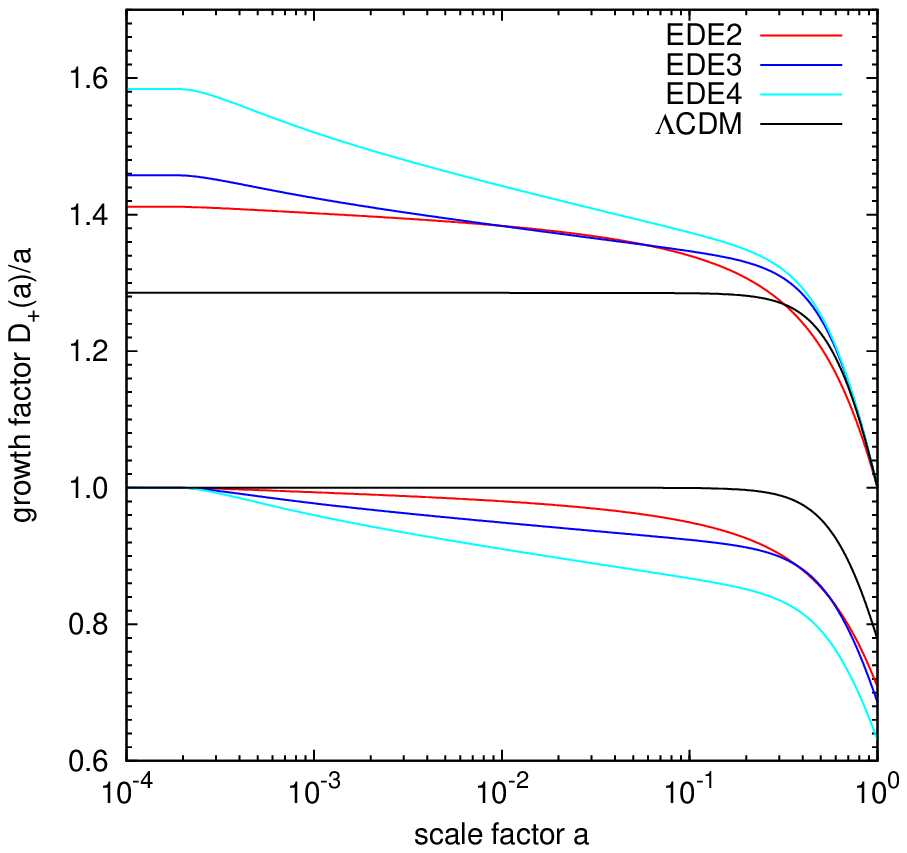}}
	\subfigure{
        	\includegraphics [height=5.6cm] {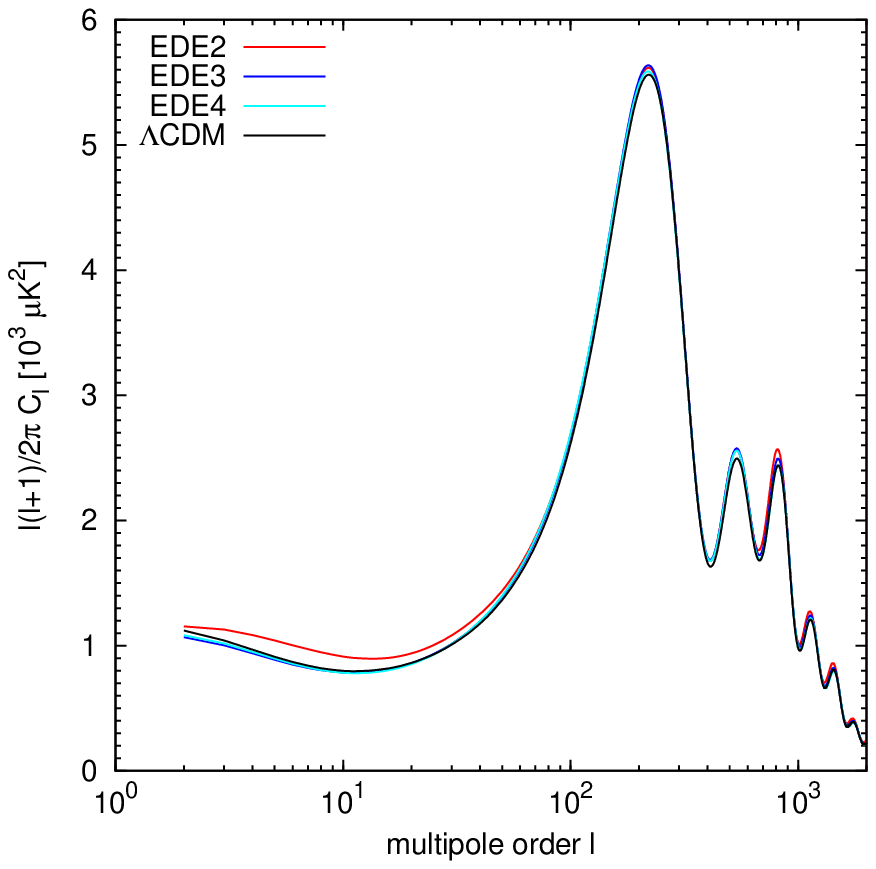}}
   \caption{Properties of the considered cosmologies. The first panel on the left shows the redshift evolution of the equation-of-state parameter \(w(z)\) (left axis) and of the \DE density \(\Omega_{\mathrm{DE}}(z)\) (right axis). The middle panel shows the evolution of the growth function with scale factor \(a\) normalised to today (upper curves) and to early times (lower curves). The panel on the right shows the CMB power spectra as computed with CMBEASY \citep{2005JCAP...10..011D}.}
   \label{Fig:EDEmisc}
\end{figure*}

\section{Early Dark Energy cosmologies}
\label{Sect2}
\subsection{General features}
Accelerated expansion in general can always be described by an equation-of-state parameter \(w_{\mathrm{DE}}=\bar{P}_{\mathrm{DE}}/\bar{\rho}_{\mathrm{DE}}\) smaller than \(-1/3\), either being constant as for example in the case of the cosmological constant \(w_{\mathrm{DE}}=-1\), or redshift depended \(w_{\mathrm{DE}}=w_{\mathrm{DE}}(z)\) as in the dynamical \DE cosmologies. The special case of a non-vanishing \DE density \(\Omega_{\mathrm{DE}}(z)\) through long periods of the cosmic evolution can be achieved by allowing that \(w_{\mathrm{DE}}(z)\) rises above zero. The left panel of Fig. \ref{Fig:EDEmisc} shows the redshift evolution of \(\Omega_{\mathrm{DE}}(z)\) and \(w_{\mathrm{DE}}(z)\) for the different cosmological models. For \(w_{\mathrm{DE}}(z)\) the tracking behaviour is clearly visible for the EDE 3 and 4 cases and less pronounced for EDE 2. The nomenclature will be introduced in Sect. \ref{subsection:2.6}. The energy densities for the EDE models do not vanish and are almost constant at \(z>10\), whereas for \(\Lambda\)CDM almost no contribution is left.
A simple parameterisation of EDE models has been introduced by \cite{2004PhLB..594...17W} and also by \cite{2006JCAP...06..026D} in which three cosmological quantities are sufficient to fully determine the model; the present \DE density \(\Omega_{\mathrm{DE},0}\), the present equation-of-state parameter \(w_{\mathrm{DE},0}\) and the averaged value of the \DE density during the era of structure formation,
\begin{equation}
	\bar{\Omega}_{\mathrm{DE,sf}}\equiv-(\ln a_{\mathrm{eq}})^{-1}\int_{\ln a_{\mathrm{eq}}}^{0}\Omega_{\mathrm{DE}}(a)\dd\ln a,
\end{equation}
where \(a_{\mathrm{eq}}\) denotes the scale factor at matter-radiation equality. Current observational constraints from SN Ia, LSS and CMB data allow for a \(\bar{\Omega}_{\mathrm{DE,sf}}\) on the percent level \citep{2005JCAP...11..007D, 2007PhRvD..75b3003D}. 

\subsection{Impact on structure formation}
The study of structure formation in the presence of \DE has been subject of many studies \citep{1998ApJ...508..483W, 2001PhRvD..64l3520D, 2003MNRAS.346..573L, 2004A&A...421...71M, 2005MNRAS.360.1393H}, leading to the result that the growth of structures is quite sensitive to the presence of \DE. In the case of EDE the non-linear part has been analysed by \cite{2006A&A...454...27B}, who showed that at the high mass end a significant enhancement of the number of clusters at redshifts \(z \sim 1\) can be expected. From the present point of view, the structures ``decay'' much slower towards the past than for the \(\Lambda\)CDM case. This could be a possible explanation for the unexpectedly high lensing efficiency of distant clusters \citep{2007A&A...461...49F} and the reported excess power at cluster scales in the CMB power spectrum \citep{2007MNRAS.380..637S}.

\subsection{The selected cosmological models}
\label{subsection:2.6}
For the present work we decided to study four cosmological models, three EDE cosmologies and the concordance \(\Lambda\)CDM model for comparison. To be consistent with preceding studies we analysed one of the two EDE models of \cite{2006A&A...454...27B} namely the EDE 2 model. In addition we used two other EDE models (3, 4) kindly provided by Georg Robbers, both being in agreement with all current bounds on EDE from SN Ia, LSS and CMB data \citep{2006JCAP...06..026D,2007PhRvD..75b3003D}, where model EDE 4 resides at the limits, giving a bordering case. The key parameters for all models can be found in Table \ref{table:1}. For all figures we chose the same colour coding to distinguish the four different models, such that red illustrates the EDE 2 model, blue and cyan depict the EDE 3 and 4 cases and grey/black denotes the fiducial \(\Lambda\)CDM case.
%
   \begin{table}
    	\caption{Key parameters of the used cosmological models.}
    	\label{table:1}
    	\centering
    	\begin{tabular}{c c c c c}
    		\hline\hline
    		Parameter & $\Lambda$CDM & EDE2 & EDE3 & EDE4\\ 
    		\hline
    		$\Omega_{\mathrm{m}}$ & 0.265 & 0.364 & 0.284 & 0.282\\
    		$\Omega_{\mathrm{DE}}$ & 0.735 & 0.636 & 0.716 & 0.718\\
    		$\bar{\Omega}_{\mathrm{DE,sf}}$ & - & 0.04 & 0.033 & 0.048\\ 
    		$h$ & 0.71 & 0.62 & 0.686 & 0.684\\
    		$\sigma_8$ & 0.8 & 0.78 & 0.715 & 0.655\\
    		$n_{\mathrm{s}}$ & 0.948 & 0.99 & 0.978 & 0.976\\
    		$w_0$ & -1 & -0.99 & -0.942 & 0.934\\
    		\hline
    	\end{tabular}
   \end{table}
%
   \begin{figure*}
   	\centering
   	\includegraphics [width=0.96\textwidth] {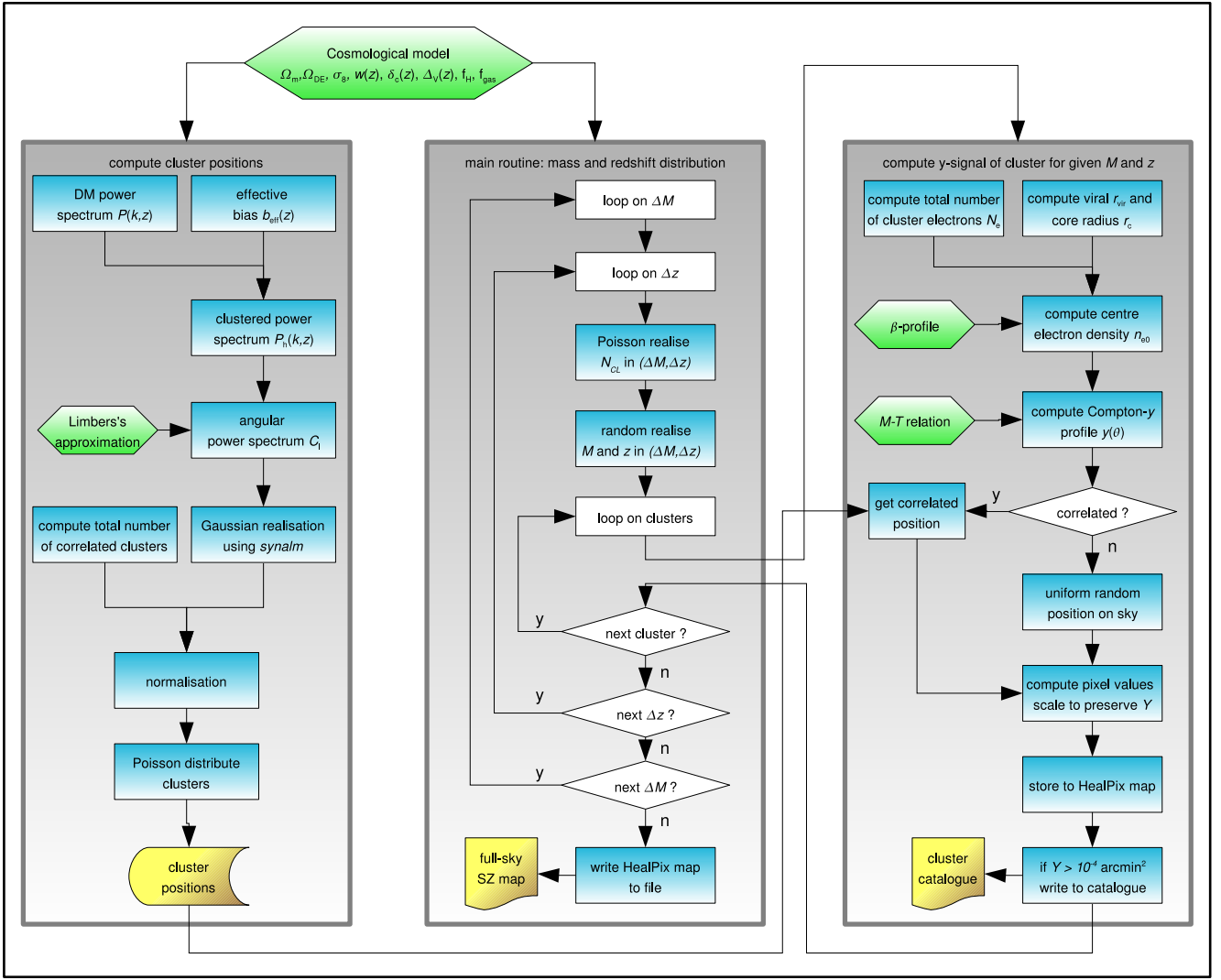}
   	\caption{Flowchart for the creation of full-sky Compton-y maps. The left column illustrates the process of obtaining the spatial cluster distribution on the sky, the centre one shows the routine for calculating the mass and redshift distributions, and the right column summarises the computation of the Compton-y profile for the individual clusters on the \healpix sky map. Green boxes illustrate physical assumptions, blue ones computations, white ones stand for algorithmic structures and yellow for results of the routines.}
    	\label{Fig:flowchart}
    \end{figure*}
%
\section{The thermal SZ effect}
The thermal SZ effect has in recent years become a popular tool for the search for clusters by current and future CMB experiments. The science case of the upcoming \planck mission also contains the task of delivering a full-sky cluster catalogue based on the SZ effect in order to study the cosmological implications.
The effect is based on the inverse Compton scattering of CMB photons off the hot electron gas in the gravitational potential of the galaxy clusters. It is identifiable by its unique spectral signature, such that at frequencies below \(\nu = 217\) GHz the clusters are observed, with respect to the CMB, as shadows and above this frequency as emitting sources. The relative temperature change \(\Delta T / T\) as a function of the dimensionless frequency \(x=h\nu/(\kB T_{\mathrm{CMB}})\) can be calculated by 
\begin{equation}
 	\frac{\Delta T}{T}(\vec{\theta})=y(\vec{\theta})\left(x\frac{e^x+1}{e^x-1}-4\right),
\end{equation}
where the amplitude, also known as the Compton-y parameter, is the line-of-sight integral of the electron pressure
\begin{equation}
	y(\vec{\theta})=\frac{\kB\sT}{m_{\e}c^2}\int n_{\e}T_{\e} \dd l. 
	\label{eq:comptony}
\end{equation}
Here \(\kB\) denotes Boltzmann's constant, \(\sT\) is the Thompson cross-section, \(m_{\e}\) is the electron mass, \(c\) denotes the speed of light, and \(n_{\e}\) and \(T_{\e}\) are the electron density and temperature, respectively. However, one is often more interested in the total Comptonisation \(\mathit{Y}\), where \(y\) is integrated over the observed cluster surface 
\begin{equation}
	\mathit{Y}  = \int \dd\Omega y(\vec{\theta}) = D_{\mathrm{A}}^{-2}(z)\frac{\kB\sT}{m_{\e} c^2}\int \dd V n_{\e}T_{\e},
\end{equation}
which can also be rewritten as an integral over the cluster volume, where \(D_{\mathrm{A}}\) denotes the angular diameter distance of the observed cluster at given redshift \(z\). For the case of isothermality the integrated Comptonisation is given by
\begin{equation}
	\mathit{Y}=\frac{\kB T_{\e}}{m_{\e}c^2}\frac{\sT}{D_{\mathrm{A}}^2}N_{\e},
\end{equation}
where \(N_{\e}\) denotes the total number of thermal electrons within the cluster, being fully determined by the cluster mass \(M_{\mathrm{cl}}\) via
\begin{equation}
	N_{\e}=\left(\frac{1+f_{\mathrm{H}}}{2m_{\p}}\right)f_{\mathrm{gas}}M_{\mathrm{cl}}
\end{equation}
for a known baryonic gas mass fraction \(f_{\mathrm{gas}}\) and hydrogen fraction \(f_{\mathrm{H}}\), where \(m_{\p}\) stands for the proton mass. For the first we assume throughout this work the cosmic value \(f_{\mathrm{gas}}=\Omega_{\mathrm{b}}/\Omega_{\mathrm{m}}=0.168\) from WMAP-5 and for \(f_{\mathrm{H}}\) the generic value \(f_{\mathrm{H}}=0.76\).

\section{Creation of full-sky SZ maps}
In order to study the impact of EDE on the detectable cluster sample, it is necessary to create full-sky maps of the thermal SZ effect. As outlined in \cite{2006MNRAS.370.1309S} one possibility is to use large scale dark matter only cosmological simulations in order to obtain the cluster positions and to additionally simulate individual clusters including gas physics to get the Compton-\(y\) signature, or to model the individual clusters by analytic \(\beta\)-profiles, as done by \cite{2005MNRAS.360...41G}. Afterwards, everything can be projected on the sphere to obtain a full-sky Compton-\(y\) map. Since for EDE cosmologies these numerical simulations are up to now not available and in order to be more flexible, we decided to take a semi-analytic approach that is outlined in the following subsections and summarised in the flowchart in Fig. \ref{Fig:flowchart}. For an overview on halo modelling in general the reader is referred to the review by \cite{2002PhR...372....1C}.
%
   \begin{figure*}
   \centering
	\subfigure{\includegraphics [height=6.5cm] {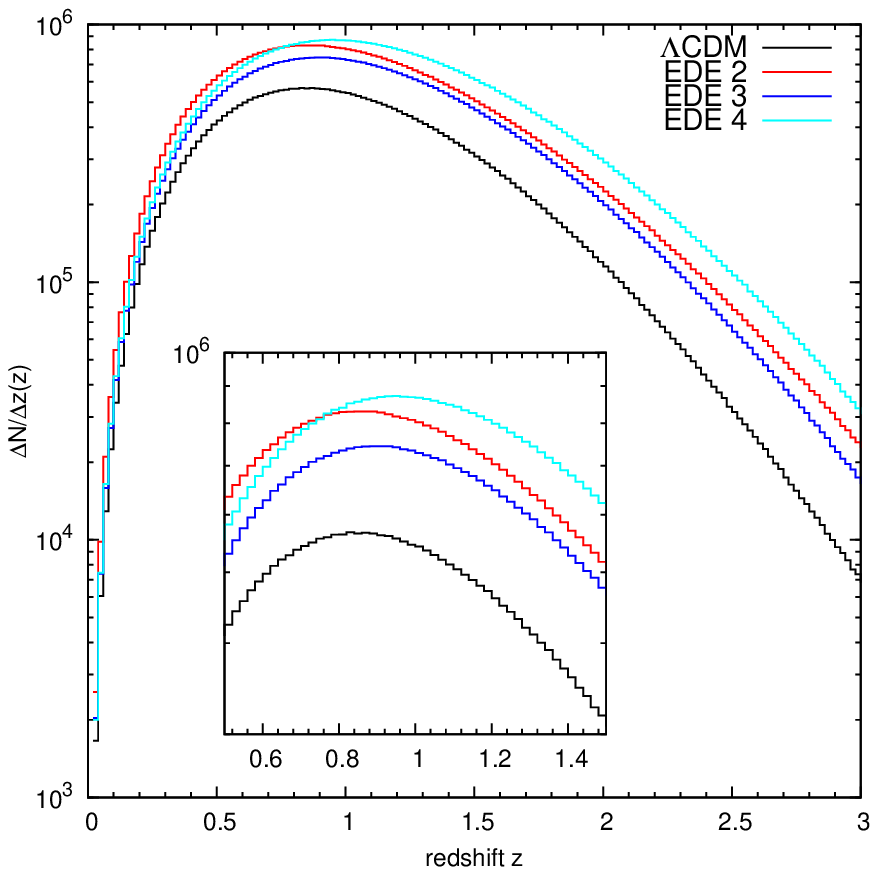}}
	\subfigure{\includegraphics [height=6.5cm] {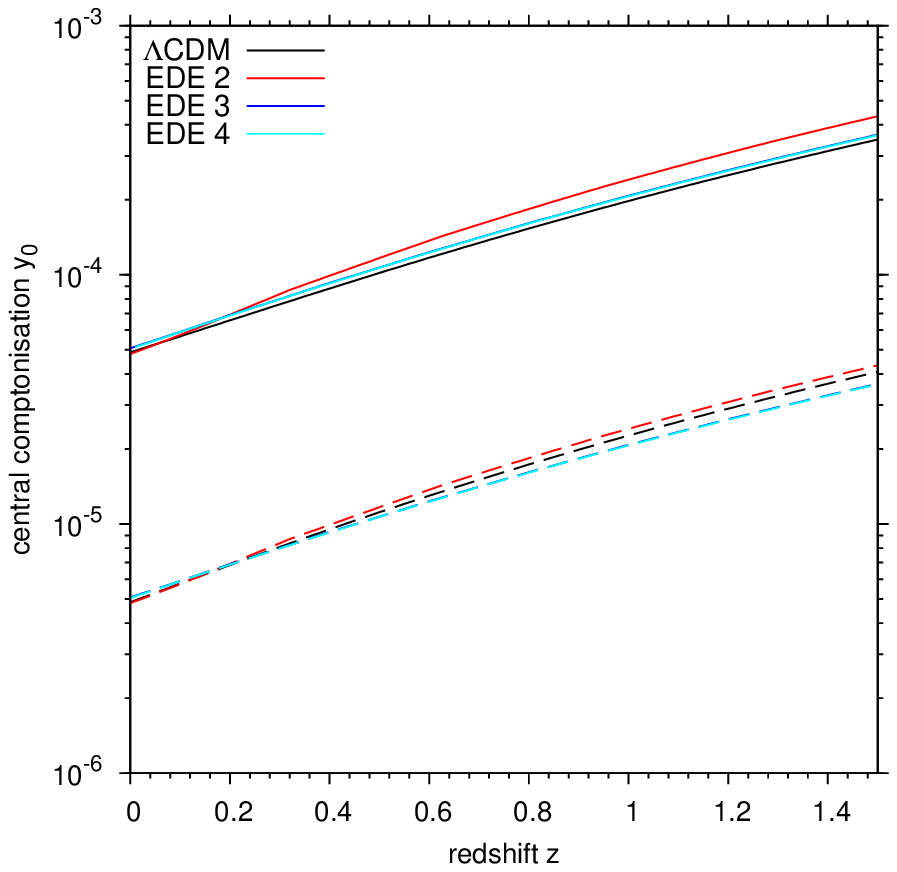}}\\
	\subfigure{\includegraphics [height=6.5cm] {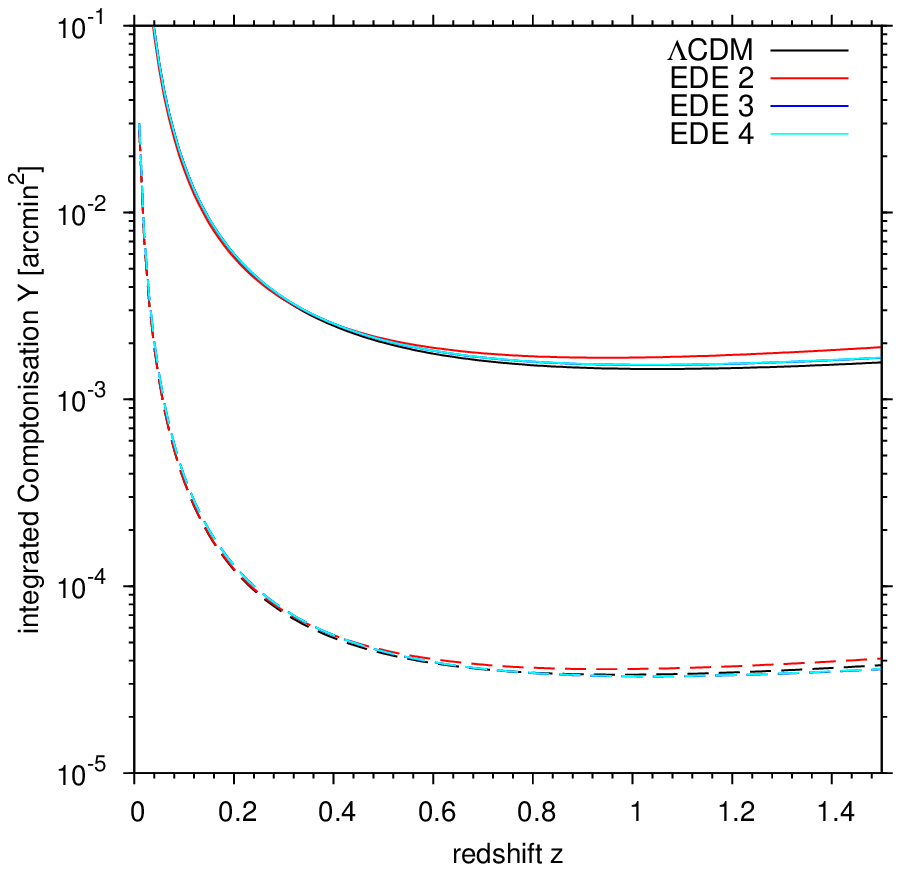}}
	\subfigure{\includegraphics [height=6.5cm] {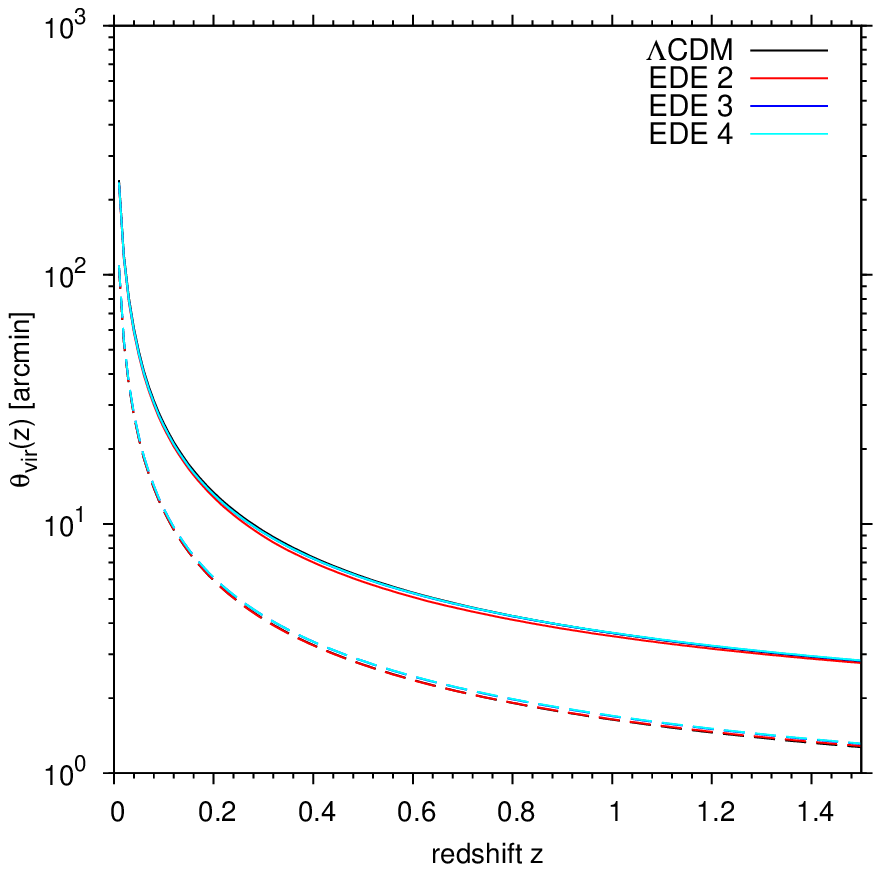}}
	\caption{Redshift evolution of the cluster properties for the \(\Lambda\)CDM and the EDE 2, 3 and 4 models. The upper left panel shows the number of Poisson-distributed clusters with \(M > 10^{13}\;h^{-1}\mbox{M}_{\odot}\) in redshift bins of \(\Delta z=0.02\) out to redshift \(z_{\mbox{max}}=3\). The other three panels show the properties of individual clusters, where solid lines denote a cluster of \(10^{15}\;h^{-1}\mbox{M}_{\odot}\) and the dashed lines are for a system with \(10^{14}\;h^{-1}\mbox{M}_{\odot}\). The upper right figure shows the central Compton-y parameter \(y_0\), the lower left the total Comptonisation \(Y\) and lower right the virial angular extent \(\theta_{\mathrm{vir}}\).} 
   \label{Fig:cluster_properties}
   \end{figure*}
%
\subsection{Collapse parameters for EDE models}
As mentioned in Sect. \ref{Sect2} the history of structure formation is quite different from the \(\Lambda\)CDM case. Basically two quantities are affected, the linear density contrast at collapse time \(\delta_c\), and the growth factor. For the calculations in the following subsections these are crucial quantities, such that we briefly summarise their computation and properties. The discussion of the spherical collapse with EDE by \cite{2006A&A...454...27B} led to the result that the critical density contrast is given by
\begin{equation}
   \delta_\mathrm{c}=\lim_{x\to0}\left[ \frac{D_+(x_\mathrm{c})}{D_+(x)}[\Delta(x)-1] \right]\;,
\end{equation}
where \(x=a/a_{\mathrm{ta}}\) is the cosmological scale factor normalised to unity at turn-around, \(D_+\) is the growth factor and \(\Delta(x)\) is the non-linear overdensity. The evolution of the growth factor for the EDE models is obtained by numerically solving the differential equation
\begin{equation}
   D_+^{\prime\prime}+\left(\frac{2}{a}+\frac{E^{\prime}}{E}\right)D_+^{\prime}-\frac{3\Omega_{\mathrm{m}}}{2a^5E^2}D_+=0,
\end{equation}
where \(E(a)\) is the expansion function of the Friedmann equation \(H^2=H_0^2E^2\), which is for the case of a time-dependend equation-of-state parameter \(w(a)\) given by
\begin{equation}
E(a)=\left\lbrace 
	\Omega_{\mathrm{m}}a^{-3}+\Omega_{\mathrm{DE}}\exp\left[ -3\int_1^a\left[1+w(a^\prime)\right]\frac{\dd a^\prime}{a^\prime}\right]
\right\rbrace^{1/2}.
\end{equation}
Therefore all cosmological quantities depending on the expansion function are modified for EDE cosmologies.\\
The virial overdensity \(\Delta_{\mathrm{V}}\) is only weakly affected by the presence of EDE as discussed in \cite{2006A&A...454...27B}, where the way of calculating \(\Delta_{\mathrm{V}}\) is explained in more detail.
\subsection{The mass and redshift distribution of the cluster sample}
The computation of the mass and redshift distribution is the core part of the map construction and is summarised in the second column of Fig. \ref{Fig:flowchart}. The basic information needed is the number of galaxy clusters per mass and redshift bin, that can be estimated using the comoving mass function
\begin{equation}
	\frac{\dd N}{\dd M \dd z} (M, z) = \Delta\Omega \frac{\dd V}{\dd z \dd \Omega} (z)
	\frac{\dd n}{\dd M} (M,z),
	\label{eq:comassfunc} 
\end{equation}
giving the cluster number density within the comoving volume element \(\dd V/\dd z \dd \Omega\) for a given solid
angle $\Delta\Omega$ on the sky. For the mass function \cite{1974ApJ...187..425P} proposed
\begin{equation}
	\frac{\dd n}{\dd M}(M,z)=\sqrt{\frac{2}{\pi}}
	\frac{\rho_0\delta_\mathrm{c}}{\sigma_R D_+(z) M^2}
	\frac{\dd\ln\sigma_R}{\dd\ln M}\nonumber \exp\left(-\frac{\delta_\mathrm{c}^2}{2\sigma_R^2D_+^2(z)}\right),
	\label{eq:massfunction}
\end{equation}
which gives the number density \(n\), of halos in \([M,M+\dd M]\) at a given redshift \(z\), where \(\rho_0\) is the present matter density, \(\sigma_R\) is the variance of the mass fluctuations and \(D_+(z)\) is the growth factor. We utilise the Press-Schechter mass function because it does not depend on numerical simulations that assume a certain cosmological model. By integrating over the mass and redshift interval of interest, one obtains the number of clusters in the aforementioned bins on the full-sky
\begin{equation}
	N_{\Delta M, \Delta z}=4\pi \int_{\Delta z}\dd z \int_{\Delta M} \dd M \frac{\dd N}{\dd M \dd z} (M, z).
 \end{equation}
The number of clusters \(N\) we assign to a bin is then obtained from a Poisson distribution with an average value of \(N_{\Delta M, \Delta z}\). In order to create the sample of the \(N\) clusters in the given bin \((\Delta M, \Delta z)\) we use a uniform random distribution. The mass bins are equally logarithmically spaced with \(\log\left(\Delta M\right)=0.1\) starting at the minimum mass \(M_{\mathrm{min}}=10^{13}\;h^{-1}\mathrm{M}_{\odot}\) up to the upper limit of \(M_{\mathrm{max}}=1\times10^{16}\;h^{-1}\mathrm{M}_{\odot}\) and the redshifts range from \(z_{\mathrm{min}}=0.01\) up to \(z_{\mathrm{max}}=3\), where the size of the redshift bin is \(\Delta z = 0.02\). Having obtained the mass and redshift distributions of the clusters, the remaining task is to model the individual SZ signal and the spatial distribution on the full-sky.

\subsection{Modelling the SZ signal of the individual clusters}

After the random realisation of \(N\) clusters in \(\Delta M\) and \(\Delta z\), both, mass and redshift, for the individual clusters are known. The next step illustrated by the right column in Fig. \ref{Fig:flowchart}, is to model the individual Compton-\(y\) signal, which can be derived from the cluster scaling relations. The mass is connected to the virial radius \(r_{\mathrm{vir}}\) by \(M=4\pi/3\Delta_{\mathrm{V}}\rho_{\mathrm{crit}}r_{\mathrm{vir}}^3\), where \(\Delta_{\mathrm{V}}\) is the mean overdensity of a virialised sphere and \(\rho_{\mathrm{crit}}\) is the critical cosmic density. Solving for \(r_{\mathrm{vir}}\) gives
\begin{equation}
	r_{\mathrm{vir}}=\frac{9.5103}{1+z}\left(\frac{\Omega_{\mathrm{m}} \Delta_{\mathrm{V}}(z)}{\Omega_{\mathrm{m}}(z)}\right)^{-1/3} \left(\frac{M_{\mathrm{cl}}}{10^{15}M_{\odot}h^{-1}}\right)^{1/3}\;h^{-1}\mbox{Mpc}.
	\label{eq:r_vir}
\end{equation}
Moreover, application of the virial theorem allows us to connect the mass to the temperature by means of
\begin{equation}
	\kB T_{\e} = \beta_{\mathrm{T}}^{-1}(1+z) \left(\frac{\Omega_{\mathrm{m}}\Delta_{\mathrm{V}}(z)}{\Omega_{\mathrm{m}}(z)}\right)^{1/3} \left(\frac{M_{\mathrm{cl}}}{10^{15}M_{\odot}h^{-1}}\right)^{2/3} \mbox{keV},
\end{equation}
with the normalisation \(\beta_{\mathrm{T}}=0.75\), being valid under the assumption of hydrostatic equilibrium and isothermality \citep[see e.g.][]{1986MNRAS.222..323K,1986RvMP...58....1S}. However, since we want to model the SZ signal for a given cluster, it is necessary to model the spatial distribution of the gas component, which is given by the \(\beta\)-profile if hydrostatic equilibrium and isothermality are assumed. Therefore the distribution of the thermal electron density is given by
\begin{equation}
	n_{\e}(r)=n_{\e 0}\left(1+\frac{r^2}{r_{\mathrm{c}}^2}\right)^{-3/2\beta},
\end{equation}
where \(r_{\cc}\) denotes the core radius. The core radius \(r_{\cc}\) and the virial radius \(r_{\mathrm{vir}}\) are related (see e.g. \cite{2005MNRAS.360...41G}) by 
\begin{equation}
	r_{\cc}(z)=\xi(z)r_{\mathrm{vir}}=0.14\left(1+z \right)^{1/5}r_{\mathrm{vir}}.
\end{equation}
Due to the lack of an according relation for EDE cosmologies we apply this relation also in these cases. In what follows we chose to set \(\beta=2/3\), delivering good fits to the X-ray surface brightness of observed clusters. \\
After these prerequisites, the first step is to fix \(n_{\e 0}\) by solving the condition \(4\pi\int_0^{r_{\mathrm{vir}}}n_{\e}(r)\, r^2\dd r=N_{\e}\) for \(n_{\e 0}\), which directly leads to
\begin{equation}
	n_{\e 0}=\frac{N_{\e}}{4 \pi r_{\mathrm{c}}^3\left[\frac{1}{\xi}+\arctan(\xi)-\frac{\pi}{2}\right] }.
\end{equation}
Since we do not integrate to infinity we are utilising a truncated \(\beta\)-profile. By using the assumption of isothermality in equation (\ref{eq:comptony}) and integrating out to \(r_{\mathrm{vir}}\) we then obtain for the central Comptonisation
\begin{equation}
	y_0=2\frac{\kB T_{\e}}{m_{\e}c^2}\sT r_{\mathrm{c}} n_{\e 0}\left[\frac{\pi}{2}-\arctan(\xi)\right].
\end{equation}
In addition, the line-of-sight integrated Compton-\(y\) profile is needed and can easily shown to be given by
\begin{equation}
	y(\theta)=2\frac{\kB T_{\e}}{m_{\e}c^2}\frac{\sT r_{\cc} n_{\e 0}}{\sqrt{1+\frac{\theta^2}{\theta_{\cc}^2}}}
	\arctan\sqrt{\frac{\frac{1}{\xi^2}-\frac{\theta^2}{\theta_{\cc}^2}}{1+\frac{\theta^2}{\theta_{\cc}^2}}},
\end{equation}
where \(\theta=r/D_{\mathrm{A}}\). In Fig. \ref{Fig:cluster_properties} we show the redshift dependence of some individual cluster properties, as well as the cluster abundance for the \(\Lambda\)CDM and EDE cases. The upper left panel clearly shows the expected increase in the number of clusters towards higher redshifts. The other three panels show individual properties of the clusters, like central and integrated Comptonisation, as well as angular extent on the sky, where all of these are only mildly affected by the presence of EDE. Therefore the main impact of EDE is the change of the abundance of galaxy clusters.

Now the machinery for calculating the signature of the individual clusters in the \healpix map is complete. First, we calculate the positions of all pixels contained in a disc of an angular size of \(\theta_{\mathrm{vir}}\), then we assign the corresponding \(y(\theta)\) calculated at the pixel centre, and in order to assure consistency with the analytic total Comptonisation \(\mathit{Y}\), we scale all pixels by a factor \(a\) given by 
\begin{equation}
	a=\frac{\mathit{Y}}{\sum_i y_i \, \dd \Omega_{\mathrm{pix}}}.
\end{equation}
Here \(y_i\) is \(y\) in the centre of pixel \(i\) and \(\Omega_{\mathrm{pix}}\) is the pixel solid angle. For clusters that are unresolved by the \healpix tessellation, we scale \(\mathit{Y}\) by the ratio of cluster solid angle to pixel solid angle. The whole procedure is performed for all clusters in the given mass and redshift range and the positions of the centre pixels are obtained as described in the following subsection.
%
   \begin{figure}
   	\centering
   	\includegraphics[width=0.9\linewidth]{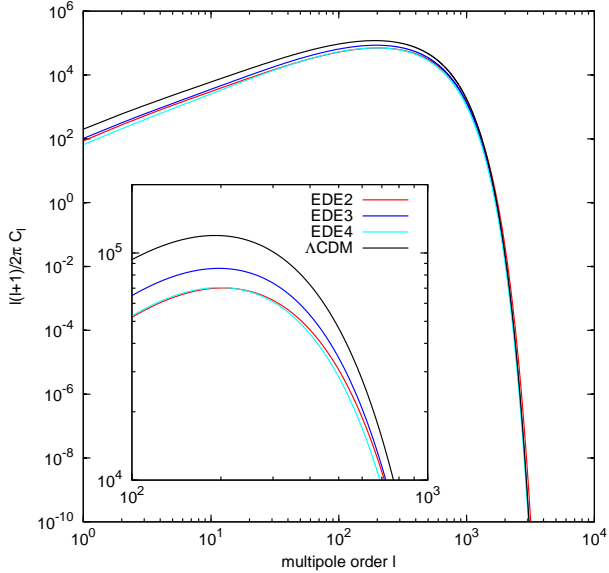}
   	\caption{Angular power spectra for the spatial distribution of clusters on the sky, assuming an averaged bias scheme for all considered cosmologies for a gaussian window function. The small panel in the plot shows a zoom in on the peak to illustrate the difference between the models.}
        \label{Fig:angular_power}
   \end{figure}
%
\subsection{Placing the clusters on the sky}
%
   \begin{figure*}
   	\centering
   	\includegraphics[width=0.9\linewidth]{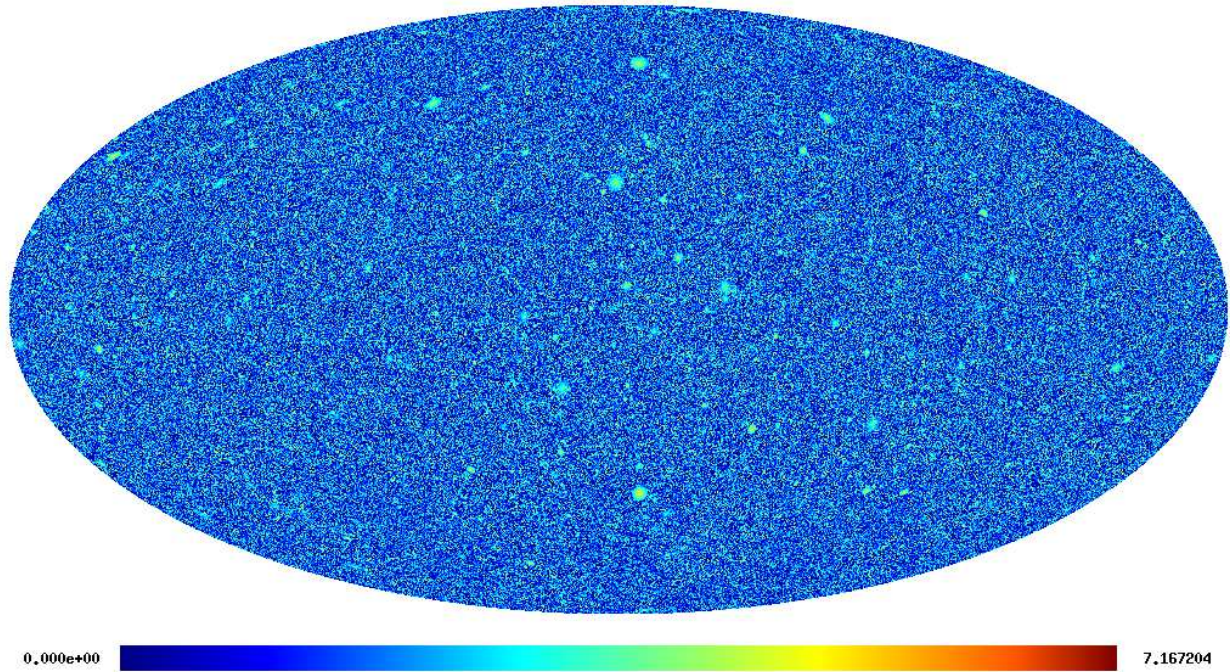}
   	\caption{Simulated full-sky SZ-map for the EDE 4 model, taking the angular correlation into account. The colour shading indicates the Compton-\(y\) parameter, being proportional to arcsinh\((10^6y)\), such that the maximum value of 7.167 corresponds to \(y=6.48\times 10^{-4}\).}
        \label{Fig:corr_scheme}
   \end{figure*}
%
After being able to model the SZ signal for clusters of given mass and redshift, the remaining task is to place them in a realistic way on the sky. Although it is quite popular to just randomly place the clusters on the sky, we decided to take the cluster correlation into account.
The scheme for obtaining the spatial distribution on the sky is presented by the left column in Fig. \ref{Fig:flowchart}, where the starting point for computing the angular correlation (or equivalently the angular power spectrum) of the clusters is the cold dark matter power spectrum given by
\begin{equation}
	P(k,z)=Ak^nT^2(k,z)D_+^2(z), 
\end{equation}
where \(D_+(z)\) denotes the growth factor, \(T(k)\) is the CDM transfer function \citep{1986ApJ...304...15B}, \(n\) is the primordial power spectrum index and the amplitude \(A\) can be normalised using \(\sigma_8\) via
\begin{equation}
	\sigma_R^2=\int_0^\infty \frac{k^2\dd k}{2\pi^2}P(k)|\tilde{W}_R(k)|^2,
\end{equation}
using the window function \(\tilde{W}_R(k)\) in \(k\)-space. Two popular choices for \(\tilde{W}_R(k)\) are the spherical top-hat window function \(\tilde{W}_R(k)=3/(kR)^3(\sin{kR}-kR\cos{kR})\) and the Gaussian smoothing \(\tilde{W}_R(k)=\exp{[-(kR)^2/2]}\). In our case both choices lead to similar results for the angular power spectra. The next step is to correct for the biasing \(b(M)\) between the halos of mass \(M\) and the dark matter distribution 
\begin{equation}
	\delta_{\mathrm{h}}=b(M)\delta_{\mathrm{dm}}, 
\end{equation}
where \(\delta_{\mathrm{h}}\) is the halo- and \(\delta_{\mathrm{dm}}\) denotes the dark matter density contrast. For the biasing we utilise the relation introduced by \cite{1996MNRAS.282..347M}, which is given by
\begin{equation}
	b(M,z)=1+\frac{1}{\delta_{\mathrm{c}}(z)}\left(\frac{\delta_{\mathrm{c}}^2(z)}{\sigma_R^2 D_+^2(z)}-1\right),
\end{equation}
introducing the mass dependence on the right hand side via the smoothing radius \(R\) in \(\sigma_R\). In order to simplify the treatment it is convenient to introduce the concept of an effective bias defined as
\begin{equation}
	b_{\mathrm{eff}}(z)=\frac{\int_{M_{\mathrm{min}}}^{M_{\mathrm{max}}} b(M,z)\frac{\dd n(M,z)}{\dd M}\dd M}{\int_{M_{\mathrm{min}}}^{M_{\mathrm{max}}}\frac{\dd n(M,z)}{\dd M}\dd M}, 
\end{equation}
which is a mass weighted integral of \(b(M,z)\) over the mass, such that we can write
\begin{equation}
	P_{\mathrm{h}}(k,z)=b_{\mathrm{eff}}^2(z)P(k,z).
\end{equation}
So far, everything is defined in 3D-space, but our goal is to obtain the angular cluster correlation with respect to the positions on the sky and to neglect correlations in redshift space. In view of this, it is natural to work with line-of-sight integrated quantities like
\begin{equation}
	\delta(\vec{n})=\int_0^{\chi_{\mathrm{max}}} \dd \chi\ q(\chi) \delta(\chi\vec{n},\chi), 
\end{equation}
which is the line-of-sight integrated density contrast in direction \(\vec{n}\) for a window function \(q(\chi)\) where \(\chi\) is the comoving distance. The expansion of \(\delta(\vec{n})\) into spherical harmonics gives 
\begin{equation}
	\delta(\vec{n})=\sum_{\ell m}\delta_{\ell m}Y_\ell^m(\vec{n}). 
\end{equation}
Putting all together, and applying Limber's approximation, one obtains for the connection between the dark matter power spectrum and the angular power spectrum the relation
\begin{equation}
	C_{\ell}^{\delta}=\int_0^{\chi_{\mathrm{max}}} \dd \chi' \frac{q^2(\chi')}{\chi'^2}P_{\mathrm{h}}\left(\frac{\ell}{\chi'},\chi'\right), 
\end{equation}
where \(q(\chi)\) denotes the window function, which is for our case a simple step function being non-zero for the intended redshift range.\\
Having obtained the angular power spectrum it is straightforward to create a Gaussian realisation using the \textit{synalm} routine of the \healpix package \citep{2005ApJ...622..759G}, representing the probability of having a cluster in a given pixel. After normalisation of the map, the number of clusters per pixel is drawn from a Poisson distribution, such that the expectation value of the cluster density equals the mean of the Poisson distribution. Figure \ref{Fig:corr_scheme} shows one full-sky Compton-\(y\) map for the case of EDE 4 with the angular correlation included. 	

\section{Simulating SZ observations by \planck}
In order to simulate realistic \planck observations one has to take several complications into account that are briefly summarised in the following subsections. First, the CMB is not the only emitting source in the frequency range of interest, our Galaxy for instance, radiates through several mechanisms like synchrotron, dust and free-free emission. Second, the detectors suffer from instrumental noise, and the scan-pattern as well as the frequency response in the different observing channels have to be taken into account. \planck will operate in nine frequency channels ranging from \(30\) GHz to \(857\) GHz with an angular resolution starting at \(33.4\) arcmin for the lowest frequency channel and going down to \(5\) arcmin for the high frequency channels above \(143\) GHz. The foreground components and the instrumental noise used for the present study are the same as in \cite{2006MNRAS.370.1309S}.

\subsection{Foreground components}
In order to perform realistic simulations of \planck observations it is necessary to include several foreground components, being active in the frequency range of interest. Following \cite{2006MNRAS.370.1713S} we assume isotropy of the spectral emission properties, as well as that the emission amplitude can be modelled by template extrapolation. For our studies we included the compulsory CMB, as well as Galactic dust, synchrotron and free-free emission:
\begin{itemize}
	\item
 	\emph{CMB:} For both the \(\Lambda\)CDM and the EDE cosmologies CMBEASY \citep{2005JCAP...10..011D} was used to generate the \(C_{\ell}\) coefficients of the CMB power spectra. From the right panel of Fig. \ref{Fig:EDEmisc} it is evident that the \(C_{\ell}\)'s for the different models are indistinguishable on cluster scales. The corresponding sets of \(a_{\ell m}\)'s have been generated using the \textit{synalm} routine.
	\item
	\emph{Synchrotron:} Being dominant at frequencies below \(100\) GHz and extending to high Galactic latitudes, the synchrotron emission is an important foreground component. The used template is based on observations \citep{1981A&A...100..209H,1982A&AS...47....1H}, being adapted to \planck observations by \cite{2002A&A...387...82G}. The spectral slope of the synchrotron emission has a spectral break at \(\nu=22\) GHz, as it has been observed by the \textit{WMAP} team.
	\item
	\emph{Dust:} For the high frequency range above \(100\) GHz Galactic dust emission is the dominant foreground component, being mostly concentrated in the Galactic disc \citep{1999ApJ...524..867F, 2000ApJ...544...81F, 1997AAS...191.8704S, 1998ApJ...500..525S}. The template derived from observations at \(\nu=3\) THz has been extrapolated to the \planck channels, using a two-component model by C. Baccigalupi, where the emission is given by the superposition of two Planck laws with \(T_1=9.4\) K and \(T_1=16.2\) K at a fixed ratio of \(0.49\).
	\item
	\emph{Free-free:} The template of the Galactic free-free emission is based on \(\mathrm{H}\alpha\)-observations \citep{2003ApJS..146..407F}, and the spectral model suggested by \cite{1998PASA...15..111V} for the conversion of the \(\mathrm{H}\alpha\) intensity to the free-free intensity. The conversion can be parameterised by the plasma temperature, leading to a \(\nu^2\)-law for the spectral dependence of the free-free brightness temperature. 
 \end{itemize}
By this, the list of contaminants in \plancks frequency range is by far not complete. In the current analysis we omitted the contribution of Zodiacal light and the emission of Solar System bodies, as well as the contamination by microwave point sources, such as star-forming galaxies and active galactic nuclei (AGN) \citep{1998MNRAS.297..117T}. Even though the impact of point sources on cluster detection is for sure not negligible, we did not attempt to model point source contamination, since the spectral behaviour, spatial clustering and biasing, as well as the AGN duty cycles, are poorly known, especially for the case of EDE cosmologies. In this sense we assume that the microwave sky has been cleaned of point source contamination prior to the cluster survey.
%
   \begin{figure}
   	\centering
   	\includegraphics[width=0.9\linewidth]{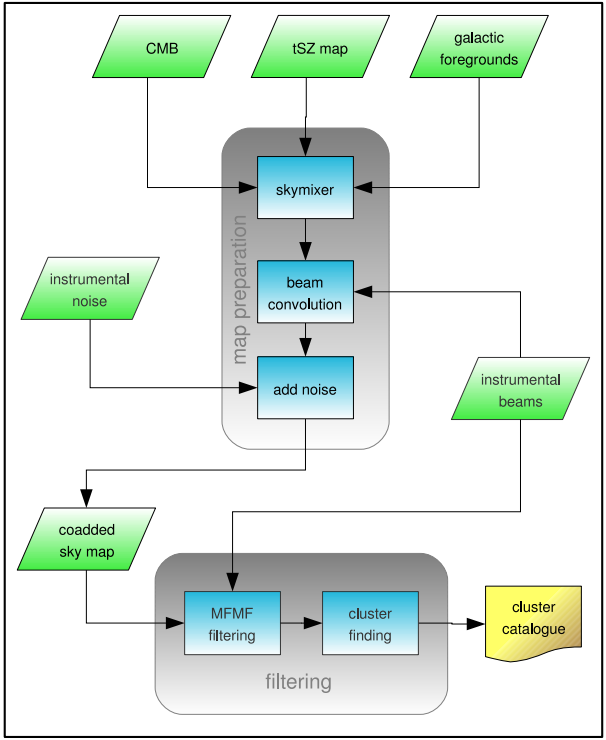}
   	\caption{Flowchart for the preparation of the coadded maps and subsequent filtering procedure.}
   	\label{Fig:flowchart_detections}
   \end{figure}
%
\subsection{Instrumental issues}
Apart from the foreground contamination, several instrumental issues have to be taken into account when modelling \planck observations, such as the angular resolution (beam size), detector noise and the frequency response of the individual channels, which are briefly summarised in the following:
\begin{itemize}
	\item 
	\planck \emph{beams:} The beam shapes for the nine channels are approximated by azimuthally symmetric Gaussians \(b(\theta)=(2\pi\sigma_{\theta})^{-1}\exp(-\theta^2/2\sigma_{\theta}^2)\) where \(\sigma_{\theta}=\Delta_{\theta}/\sqrt{8\ln(2)}\) with \(\Delta_{\theta}\) being the angular resolution, starting from \(33.4\) arcmin for the lowest channel and going down to \(5.0\) arcmin for all channels above \(143\) GHz. 
	\item 
	\emph{Noise maps:} The noise per pixel is obtained by drawing a Gaussian distributed random number with zero mean and and width \(\sigma_{\mathrm{N}}\) in units of antenna temperature, which is obtained from the \planck detector database. Assuming Poisson statistics, the noise level can be scaled down by \(\sqrt{N_{\mathrm{det}}}\) with \(N_{\mathrm{det}}\) being the number of redundant detectors for each channel, and by \(\sqrt{N_{\mathrm{hit}}}\), which is the number of observations in the pixel of interest.
	\item 
	\emph{Frequency response:} \plancks frequency response can be well approximated by a top-hat, centred on the fiducial centre frequency \(\nu_0\), with a channel dependent window size \(\Delta\nu\).
\end{itemize}

\subsection{Preparation of coadded maps}
The preparation of the observed coadded sky maps, including all astrophysical components and instrumental properties, is presented in the upper part of the flowchart shown in Fig. \ref{Fig:flowchart_detections}. First, the {\em skymixer} module, being part of the \planck LevelS pipeline \citep{2006A&A...445..373R}, is used to add up the templates of the CMB, the galactic foreground components and the SZ effect in \(a_{\ell m}\)-space, applying the spectral laws for each component for the given frequency response in the respective channel. After that, the maps are convolved with the beam profiles of the respective channel and the noise maps are added, leaving nine coadded maps \(\mathit{C}_{\ell m}\) in \(a_{\ell m}\)-space that can directly be fed into the filtering routines as described in the next section. 
%
   \begin{figure*}
   	\centering
	\subfigure{\includegraphics [height=5.5cm] {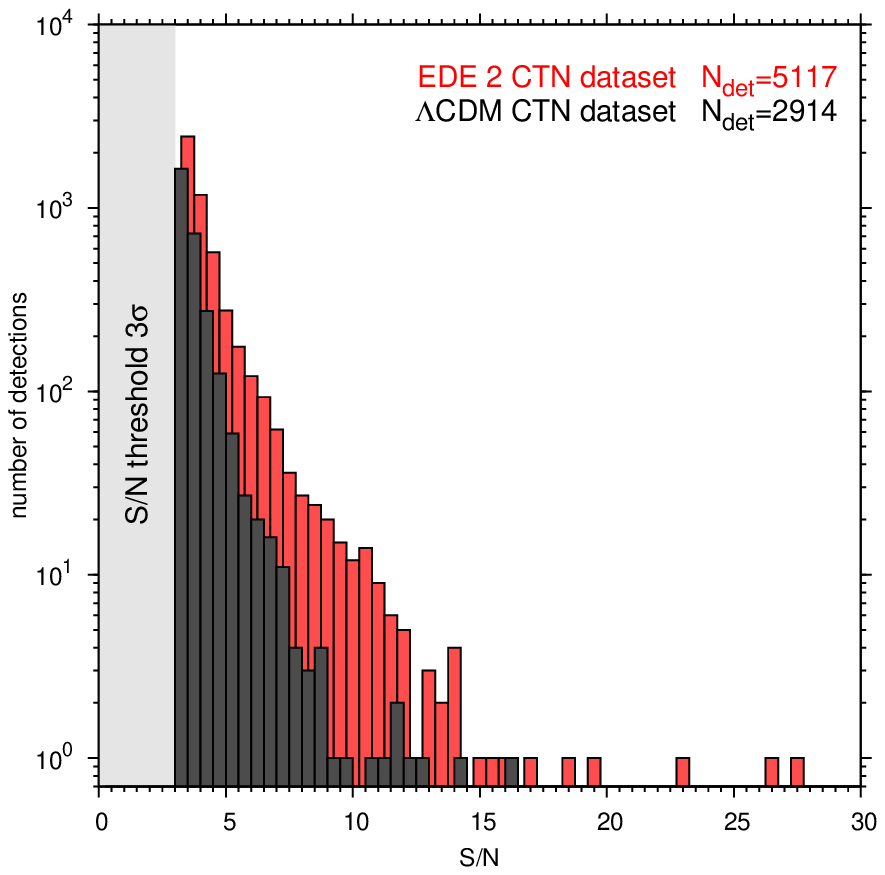}}
	\subfigure{\includegraphics [height=5.5cm] {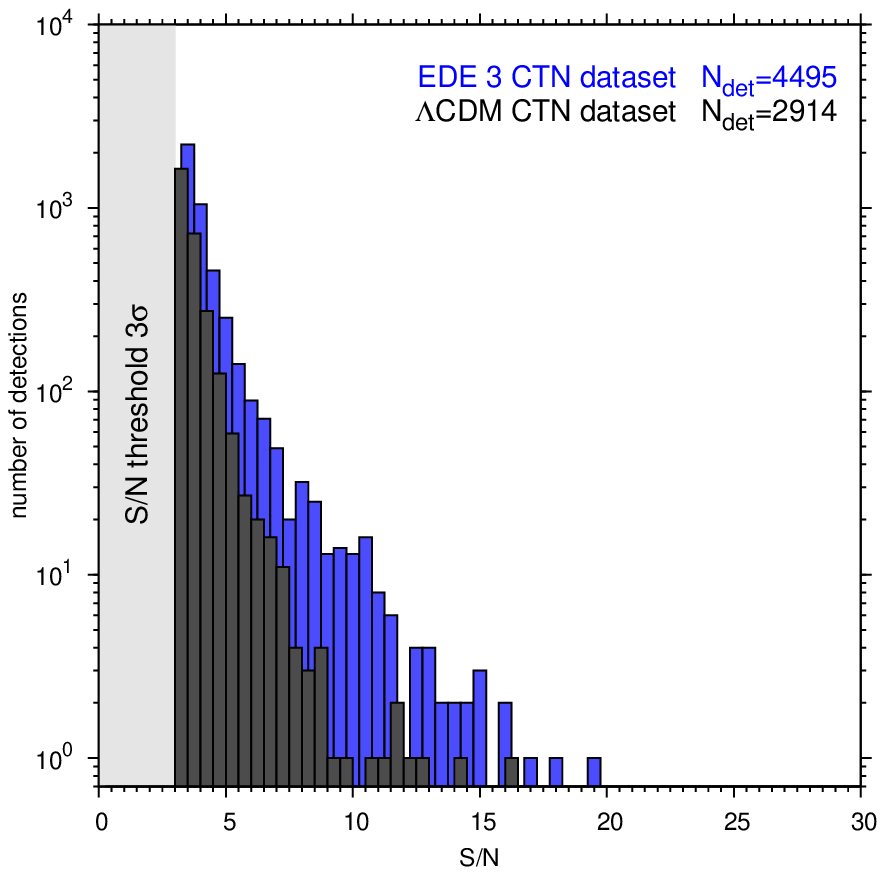}}
	\subfigure{\includegraphics [height=5.5cm] {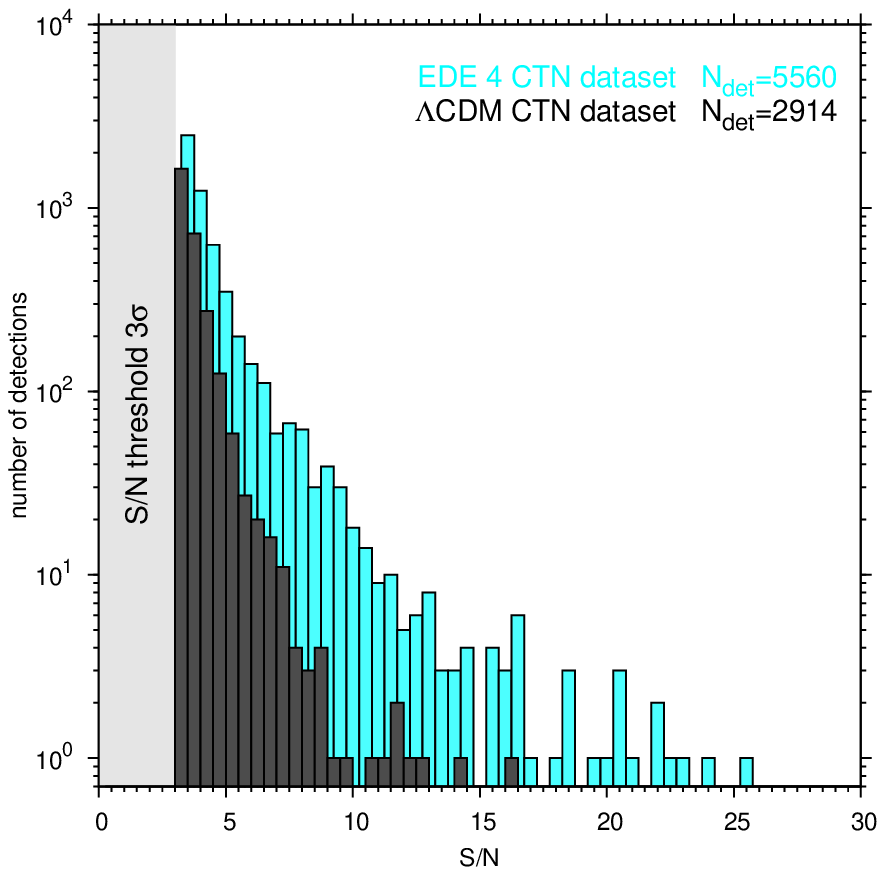}}\\
	\subfigure{\includegraphics [height=5.5cm] {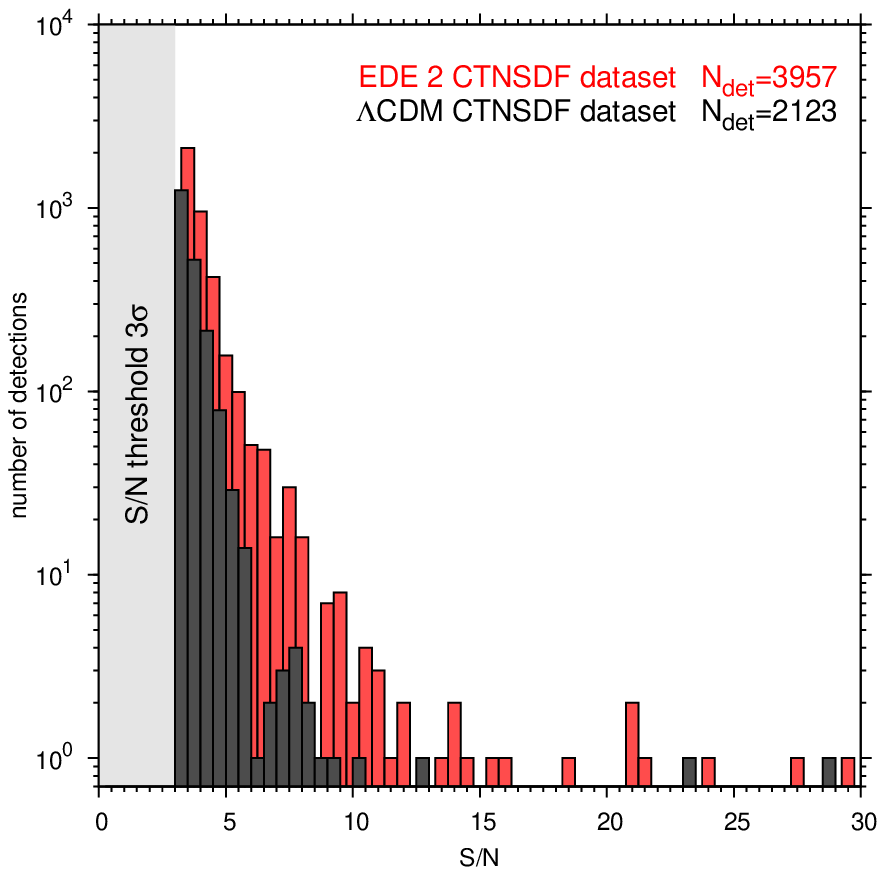}}
	\subfigure{\includegraphics [height=5.5cm] {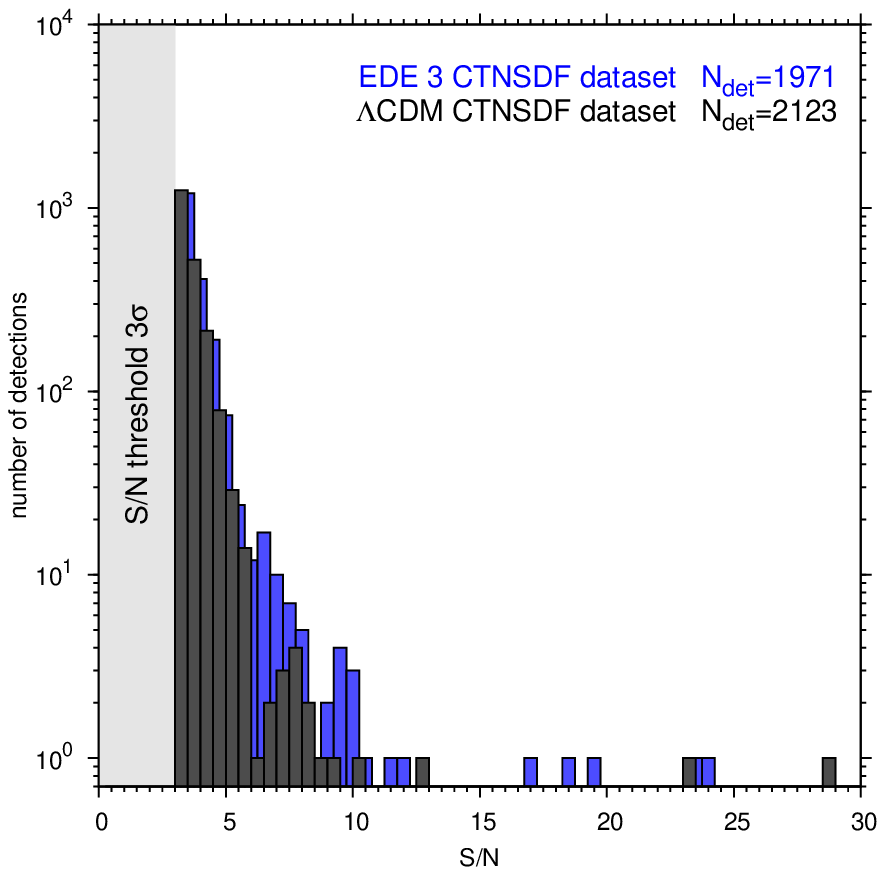}}
	\subfigure{\includegraphics [height=5.5cm] {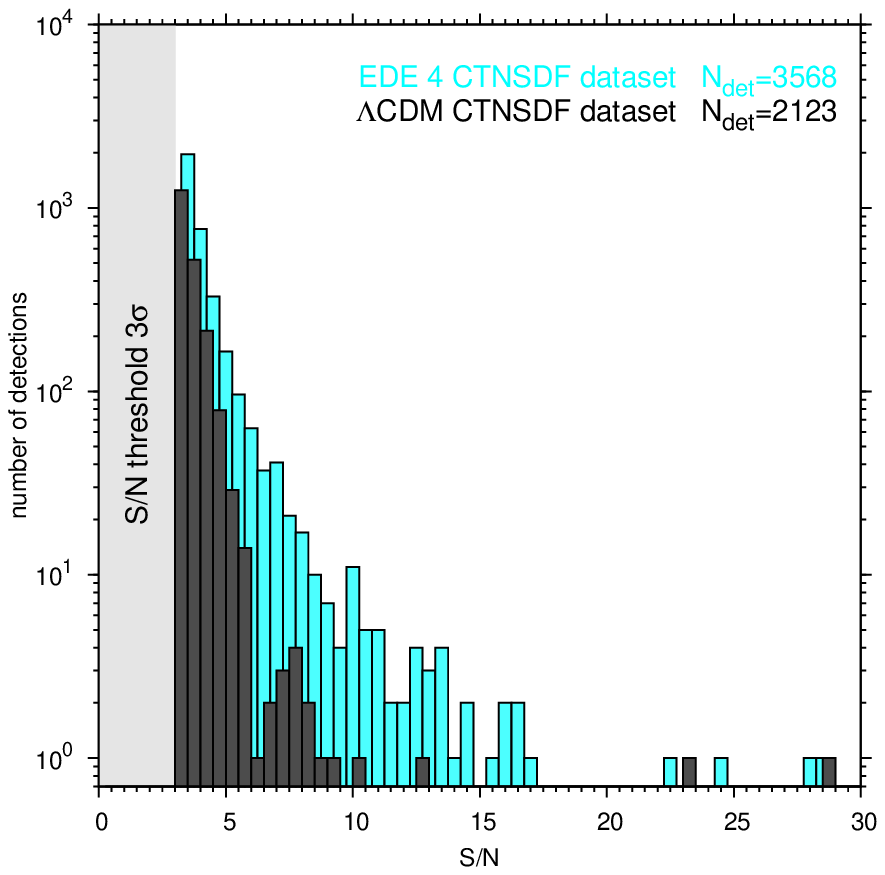}}
	\caption{Number of confirmed detections in bins of S/N of width \(\Delta_{\mbox{S/N}}=0.5\) for \(\mathit{Y}_{\mathrm{treshold}}=2\times10^{-4} \; \mathrm{arcmin}^2\) and an association radius of \(r_{\mathrm{search}}=30\;\mathrm{arcmin}\) for the EDE 2, EDE 3 and EDE 4 models from left to right. The upper row shows the clean datasets (CTN) and the lower one the foreground contaminated datasets (CTNSDF). For comparison also the \(\Lambda\)CDM case is also shown in each plot. The grey strip on the left indicates the filtering detection threshold of \(3\,\sigma\).}
   	\label{Fig:sn_ctn}
    \end{figure*}
%
\section{The filtering}
In order to study the impact of EDE on the observed cluster sample we decided to use the multi-frequency-matched-filter (MFMF) method introduced by \citep{2006MNRAS.370.1713S, 2007MNRAS.377.253S}. It is a generalisation of the matched filtering, as introduced by \cite{2001ApJ...552..484S} and \cite{2002MNRAS.336.1057H}, to spherical datasets formulated in spherical harmonic space. The derivation of the filter kernels \(\psi(|\theta|,R)\), being constructed for the detection of objects of a certain scale \(R\), can be formulated as the solution of a variational problem. The detection map \(\mathit{D}_i(\theta,R)\) for channel \(i\) is obtained by convolving the corresponding filter kernel \(\psi_i(|\theta|,R)\) with the coadded map \(\mathit{C}_i(\theta)\), where the variance \(\sigma_{\mathit{D}}^2(R)\) of the detection map is constrained to be minimal. Additionally the filter kernels have to fulfil two further conditions:
\begin{enumerate}
 	\item to ensure the existence of a scale \(R_0\), such that the filtered field \(\langle\mathit{D}(\theta_{\mathrm{source}},R_0)\rangle\) at the source position \(\theta_{\mathrm{source}}\) is maximal and that
 	\item the filter kernel should be an unbiased estimator of the underlying source amplitude. 
\end{enumerate}
The source profiles needed for the filter construction are assumed to be spherically symmetric \(\beta\)-profiles and to be superimposed on a homogeneous and isotropic fluctuating background, obeying Gaussian statistics. In this sense it is further assumed, that all foregrounds have isotropic spectral properties. Further detail on the filter construction, as well as the results of extensive testing of the filter performance and the key assumptions made, can be found in \citep{2006MNRAS.370.1713S, 2007MNRAS.377.253S}.

For our case the filtering procedure is summarised in the flowchart shown in Fig. \ref{Fig:flowchart_detections}. After the construction of the coadded Compton-\(y\) maps in \(a_{\ell m}\)-space, as shown in the upper part of the flowchart, filters optimised for finding beam convolved \(\beta\)-profiles are constructed and normalised by the variance of the coadded sky maps. In doing so, the convolution of the filters with the coadded maps and back-transforming to real-space, yields a detection map, giving a signal-to-noise ratio (S/N) for each pixel. Applying a S/N threshold of \(3\,\sigma\) and a Galactic cut of \(10\,\deg\), prepares the detection map for the last step of cluster identification. In this step all peaks above the threshold are identified and the position in Galactic coordinates, as well as the peak S/N are dumped to the cluster catalogue, which can afterwards be cross-checked against the input cluster catalogue. 
The fact that the method works on the full-sky, is fast and blind makes it an ideal tool for our purposes in allowing fast parameter studies. For a given set of coadded maps it take less than 45 minutes on a normal PC to obtain a cluster catalogue.
%
  \begin{table}
  	\caption{Properties of the simulated maps, \(N_{\mbox{corr}}\) is number of correlated clusters and \(N_{\mbox{tot}}\) is the total number of simulated systems.}
  	\label{table:2}
  	\centering
  	\begin{tabular}{c c c c c}
  		\hline\hline
  		map & $N_{\mbox{corr}}$ & $N_{\mbox{tot}}$\\ 
  		\hline
  		$\Lambda$CDM & $1.02\times 10^5$ & $3.58\times 10^7$\\
  		EDE 2 & $3.08\times 10^5$ & $5.68\times 10^7$\\
  		EDE 3 & $2.46\times 10^5$ & $5.02\times 10^7$\\ 
  		EDE 4 & $4.10\times 10^5$ & $6.21\times 10^7$\\
  		\hline
  	\end{tabular}
  \end{table}
%
\section{Results}
To perform our study we created four full-sky Compton-\(y\) maps; three for the EDE 2, 3 and 4 models and one for the fiducial \(\Lambda\)CDM cosmology. The maps contain a cluster sample with masses ranging between \(10^{13}\;h^{-1}\mbox{M}_{\odot}\) and a few times \(10^{15}\;h^{-1}\mbox{M}_{\odot}\) out to redshifts of \(z_{\mathrm{max}}=3.0\). In order to speed up the computation, the angular cluster correlations have been taken into account for systems with \(M_{\mathrm{cl}}\geq 1.5\times10^{14}\;h^{-1}\mbox{M}_{\odot}\), such that we assume a homogeneous and isotropic SZ background. This threshold has been chosen in order to ensure numerical feasibility, as well as minimising the effect on the detected cluster sample. Varying the threshold actually showed, that the statistical impact on the detected cluster sample for our filter method is negligible. Depending on the model, the maps comprise in total more than sixty million galaxy clusters as summarised in Table \ref{table:2}, ensuring a sufficient treatment of the SZ-background due to unresolved clusters. We fed these maps into our observation and filtering pipelines for the clean case, where we just added the CMB and instrumental noise to the maps, dubbed hereafter the CTN case, and for the CTNSDF case, where we added also the Galactic foregrounds.\\
Assuming a survey threshold of \(\mathit{Y}_{\mathrm{treshold}}=2\times 10^{-4}\;\mathrm{arcmin}^2\), corresponding to a limiting mass \(M_{min}=3.5\times10^{14}\;h^{-1}\mbox{M}_{\odot}\) at \(z=1.0\), and an association radius of \(r_{\mathrm{search}}=30\;\mathrm{arcmin}\) we cross-check against the input cluster catalogues of our Compton-\(y\) maps and obtain in this way cluster catalogues that are free of false detections. In this sense we assume a \planck cluster catalogue comprising only confirmed detections, where the validity and redshift of each detection stem from follow-up observations. Moreover, our study does not aim to recover individual cluster properties, like mass or the integrated Comptonisation, but to exploit the effect on the detected number counts.\\   

\begin{figure*}
   	\centering
	\subfigure{\includegraphics [height=5.5cm] {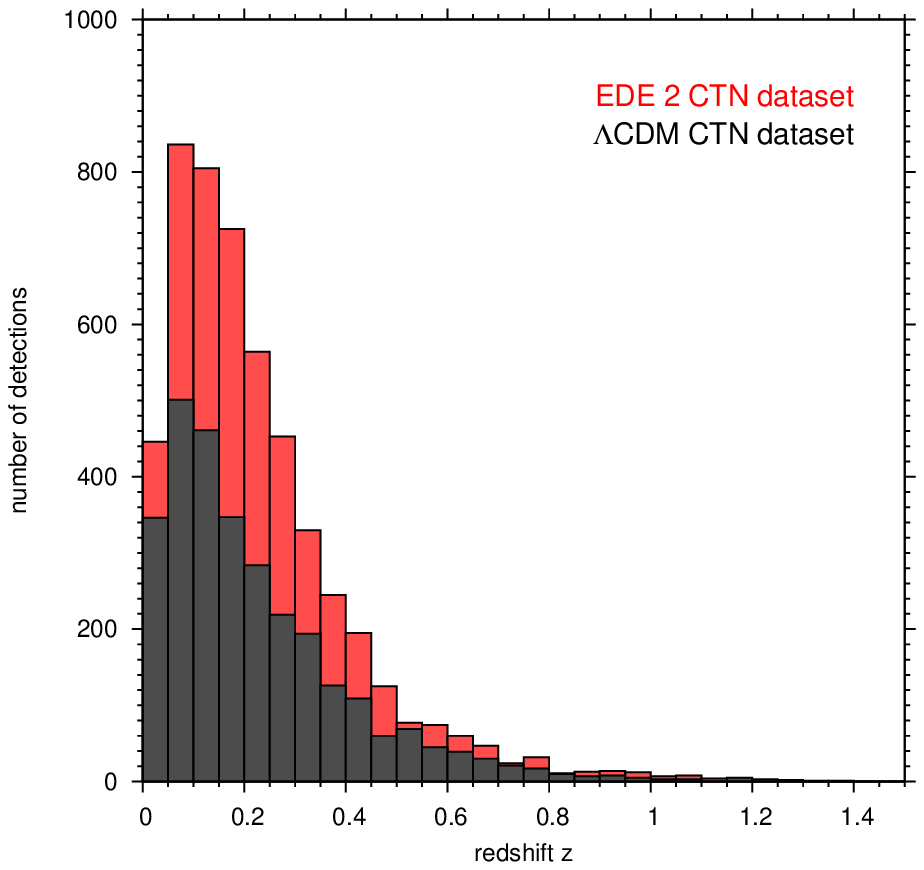}}
	\subfigure{\includegraphics [height=5.5cm] {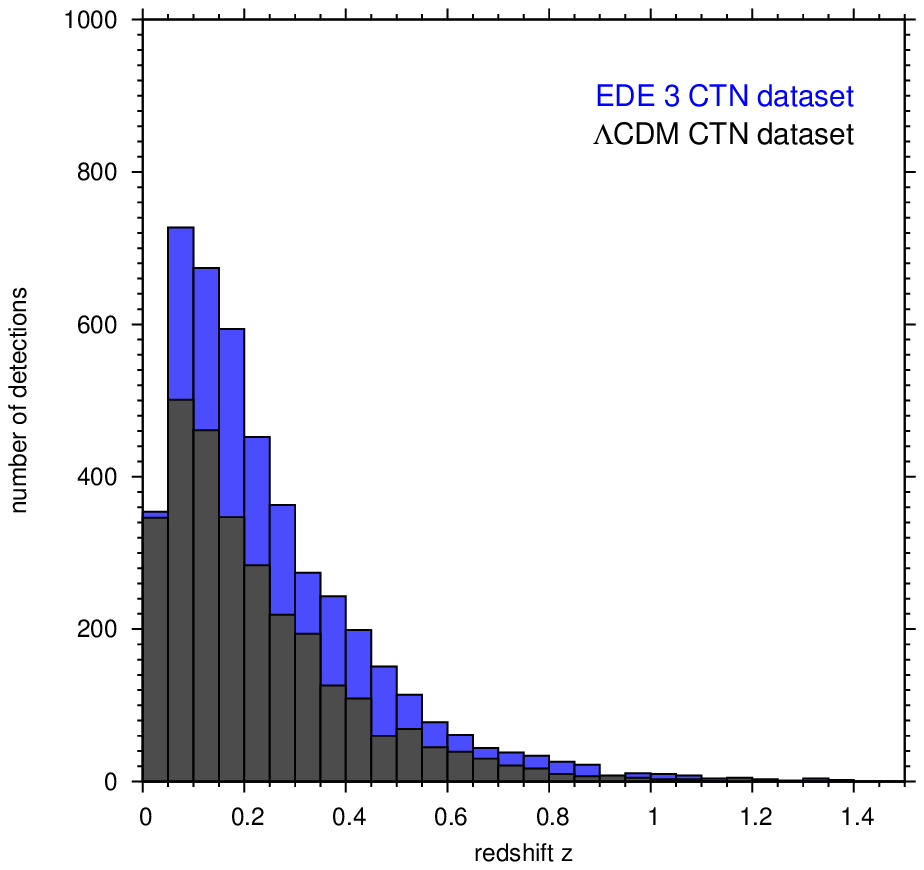}}
	\subfigure{\includegraphics [height=5.5cm] {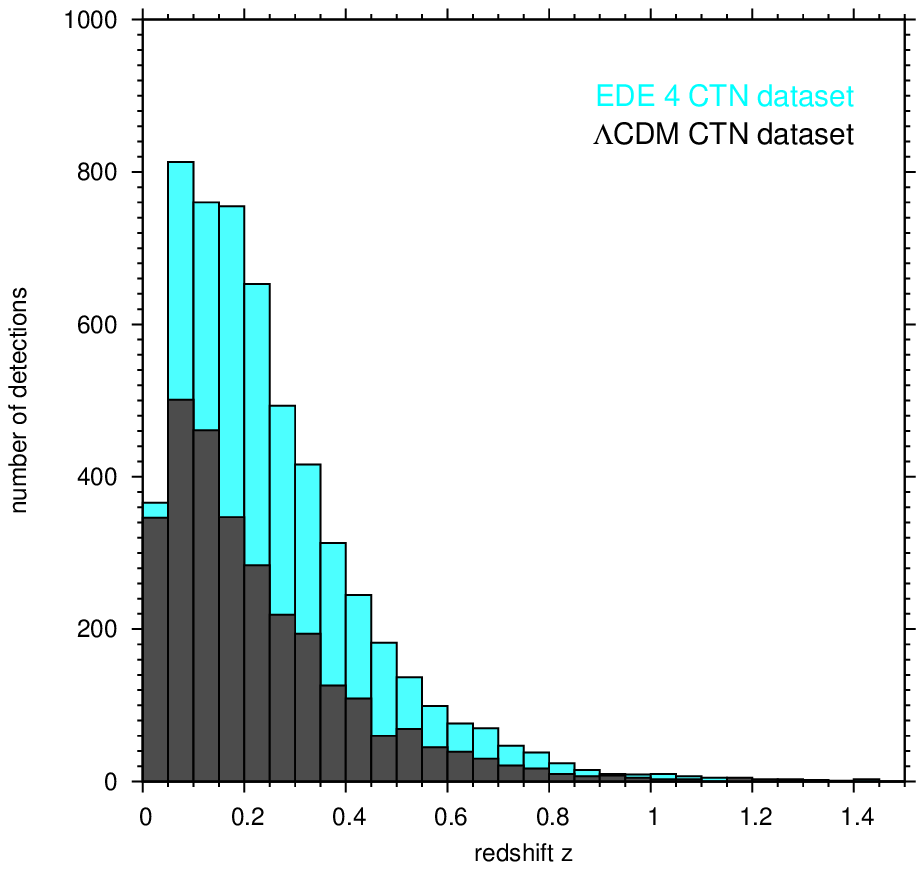}}\\
	\subfigure{\includegraphics [height=5.5cm] {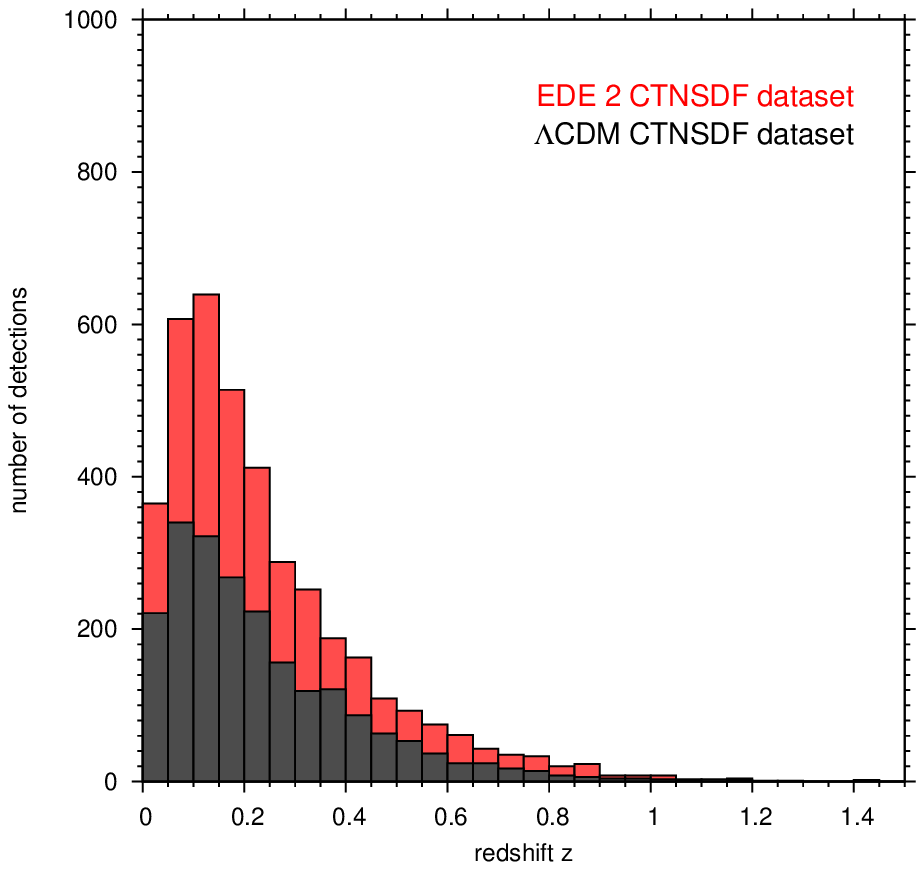}}
	\subfigure{\includegraphics [height=5.5cm] {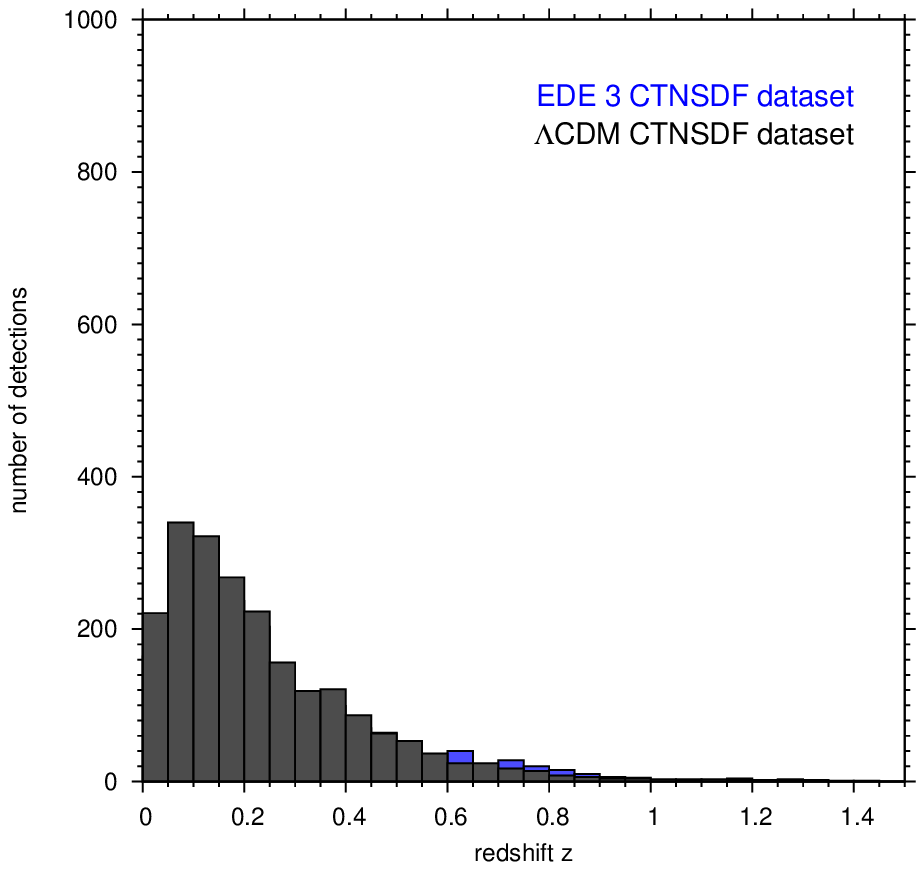}}
	\subfigure{\includegraphics [height=5.5cm] {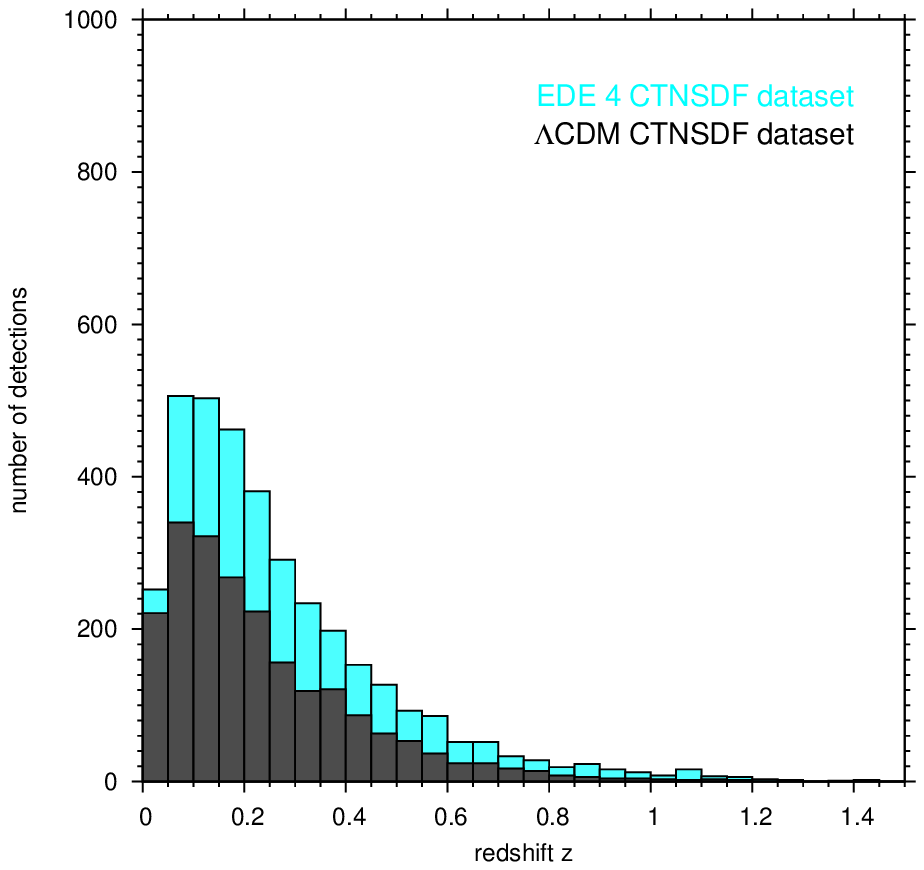}}
	\caption{Number of detections in redshift bins of width \(\Delta z=0.05\) for \(\mathit{Y}_{\mathrm{treshold}}=2\times10^{-4} \; \mathrm{arcmin}^2\) and an association radius of \(r_{\mathrm{search}}=30\;\mathrm{arcmin}\) comprising the EDE 2, EDE 3 and EDE 4 models from left to right for the clean case (CTN) in the upper row and for the galactic foreground dataset (CTNSDF) in the lower one. To ease comparison the \(\Lambda\)CDM case is also shown in each plot as well.}
   	\label{Fig:sn_z_bins}
    \end{figure*}

The first step is to look at the distribution of the number of detections in bins of S/N, as shown in Fig. \ref{Fig:sn_ctn}, which is the information directly obtained from the filtering process. For the clean CTN case (upper row) as well as for the CTNSDF case (lower row), the curves in Fig. \ref{Fig:sn_ctn} look as expected, in the sense that the number of detections is a decreasing function of the S/N. However, the deviations between the \(\Lambda\)CDM case and the EDE cases are evident. For S/N above \(5\sigma\) the number of detections is much higher (up to \(100\%\) and more) in some bins, and overall the distribution falls of flatter with the S/N. This result is not surprising since we expect in EDE cosmologies a higher abundance of massive systems in the past and these systems are quite likely giving rise to higher S/N. However, when having a closer look on the EDE 3 CTNSDF case in Fig. \ref{Fig:sn_ctn}, it is obvious that for this case the difference is less pronounced and can hardly be distinguished from the \(\Lambda\)CDM case. \\
After the assumed redshift determination by follow-up observations, it is also possible to study the distribution of detected systems in redshift bins. The results for a redshift bin size of \(\Delta z=0.05\) are shown in Fig. \ref{Fig:sn_z_bins}, again for all models, and the CTN as well as the CTNSDF datasets. In the redshift range \(0<z<0.6\) the increase in number of detections for EDE relative to \(\Lambda\)CDM is quite pronounced for the clean case and also for the EDE 2 and 4 CTNSDF datasets, whereas the EDE 3 CTNSDF case is indistinguishable from a \(\Lambda\)CDM model in terms of the redshift distribution. As expected the strongest detectability is found for the limiting EDE 4 case, resulting in a boosted number of detections up to \(z\sim1\). When taking the values of \(\bar{\Omega}_{\mathrm{DE,sf}}\) from Table \ref{table:1} into consideration, one can argue that with the used detection procedure EDE models having around \(4\%\) of \(\bar{\Omega}_{\mathrm{DE,sf}}\) should be detectable.\\

In addition to the number counts, EDE has an effect on the contamination of the detected cluster sample. We define the contamination as the ratio of the number of true detected \(N_{\mbox{true}}\) clusters over the total number of claimed detections \(N_{\mbox{det}}\). The values for all four setups can be found in Table \ref{table:3} from which it is evident that all EDE catalogues are significantly less contaminated than the \(\Lambda\)CDM case. Such an effect can be explained by the impact of the relative noise levels for the \(\Lambda\)CDM and EDE cases, such that for the EDE case one has more significantly detectable clusters relative to the uniform background than for \(\Lambda\)CDM. This effect can also be mimicked by downscaling the noise as shown in the appendix. In this sense EDE helps its own detectability not only by increasing the number counts, but also by improving the purity of the detected cluster sample.\\
 
%
\begin{figure}
   	\centering
   	\subfigure{\includegraphics[width=0.49\linewidth]{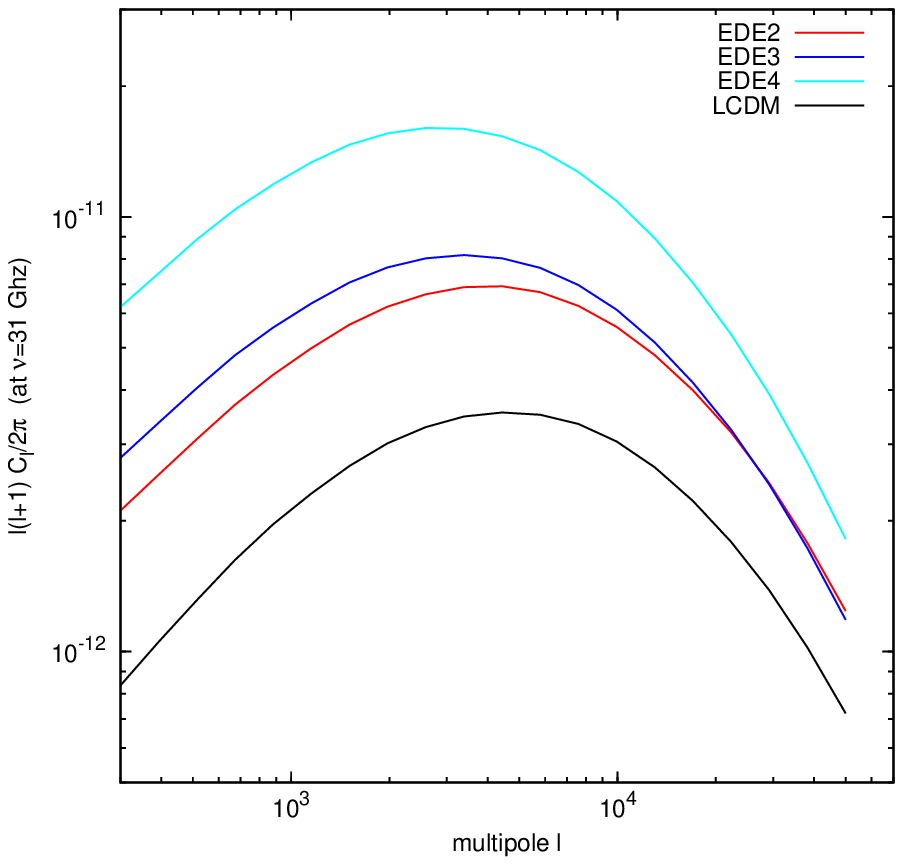}}
   	\subfigure{\includegraphics[width=0.49\linewidth]{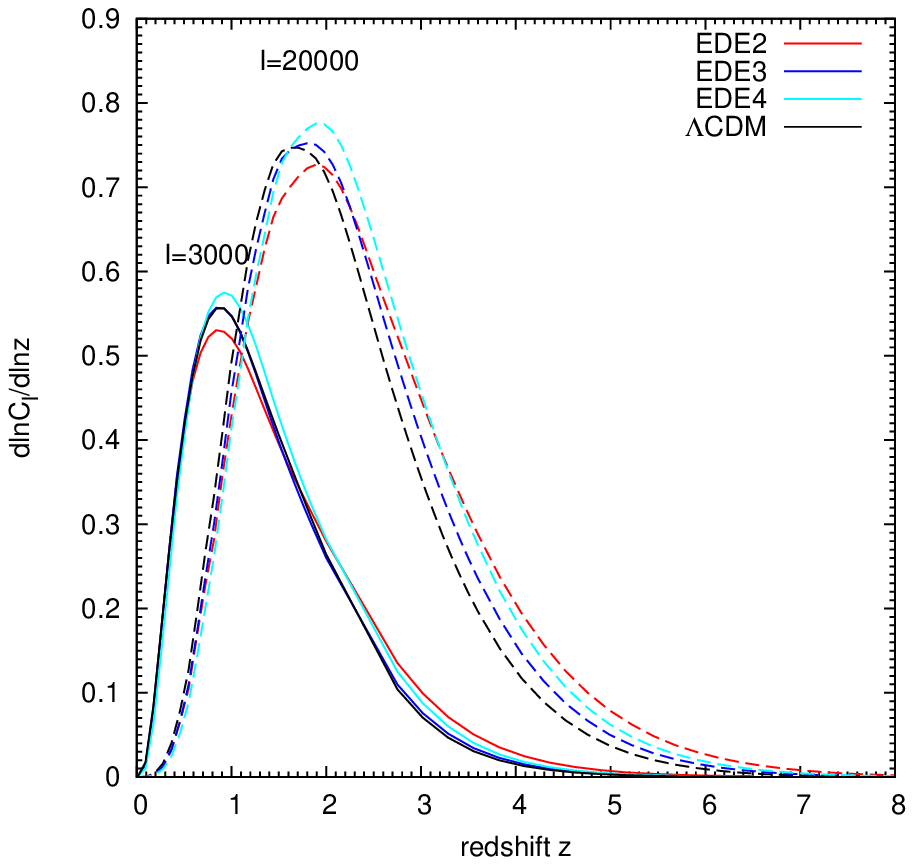}}
   	\caption{SZ angular power spectra at \(\nu = 31\;\mathrm{GHz}\) (left panel) and redshift distribution of \(C_{\ell}\) (right panel), plotted as \(\dd \,\ln C_{\ell}/\dd\, \ln z\) for \(\ell=3000,20000\), for the EDE2, EDE3 and EDE4 as well as the fiducial \(\Lambda\)CDM model.}
        \label{Fig:SZ_power_spectra}
   \end{figure}

%
  \begin{table}
  	\caption{Properties of the detected cluster sample for \(\mathit{Y}_{\mathrm{treshold}}=2\times10^{-4} \; \mathrm{arcmin}^2\) and an association radius of \(r_{\mathrm{search}}=30\;\mathrm{arcmin}\), \(N_{\mbox{det}}\) is number of detections clusters and \(N_{\mbox{true}}\) are the true ones.}
  	\label{table:3}
  	\centering 
  	\begin{tabular}{c c c c c}
  		\hline\hline
  		map & $N_{\mbox{det}}$ & $N_{\mbox{true}}$ & contamination [\%] \\  
  		\hline
  		$\Lambda$CDM CTN& $4221$ & $2914$ & 30.96\\
  		EDE 2  CTN & $5272$ & $5117$ & 2.94\\
  		EDE 3  CTN & $4796$ & $4495$ & 6.28\\ 
  		EDE 4  CTN & $5651$ & $5560$ & 1.61\\
  		\hline 
  		$\Lambda$CDM CTNSDF& $3389$ & $2123$ & 37.36\\
  		EDE 2  CTNSDF & $4124$ & $3957$ & 4.05\\
  		EDE 3  CTNSDF & $2142$ & $1971$ & 8.41\\ 
  		EDE 4  CTNSDF & $3680$ & $3568$ & 3.04\\
  		\hline
  	\end{tabular}
  \end{table}
In the introduction it was mentioned that EDE also might offer an explanation for the excess in angular power at high multipoles, as discussed by \cite{2007MNRAS.380..637S}. To show that this is also the case for our choice of models we calculated the SZ power spectra following the approach as discussed, for example, by \cite{2000PhRvD..62j3506C}, \cite{2002MNRAS.336.1256K} and \cite{2002PhRvD..66d3002R}. In order to be comparable to \cite{2007MNRAS.380..637S} we show in Fig.\ref{Fig:SZ_power_spectra} the SZ power spectra at an observing frequency of \(\nu = 31\;\mathrm{GHz}\) and the redshift distribution of \(C_{\ell}\). As expected for the power spectra, all EDE models lie above our fiducial \(\Lambda\)CDM cosmology and for the EDE4 model the boost in power is strongest, despite having the lowest \(\sigma_{8}=0.655\), showing that EDE could in principle offer an explanation to excess power at high multipoles. The redshift distribution of the \(C_{\ell}\) clearly shows an enhancement in the high redshift tail, since EDE contributes strongest at higher redshifts. Even though the current observational status does not allow the models to be distinguished, and the nature of the excess is still unclear, future high quality measurements of the SZ power spectra will be a powerful tool for confirming, or ruling out, deviations from the \(\Lambda\)CDM model.
%
\section{Summary and conclusions}
The aim of this paper is to present the results of a study on the impact of the altered structure formation history in EDE cosmologies on \plancks SZ cluster sample, using the spherical MFMF filtering approach. By computing the SZ power spectra, we show that EDE indeed leads to a boost in the level of power, which could in principle explain the excess of power on small angular scales of the observed CMB power spectrum.\\
In order to examine the impact of EDE on \plancks SZ cluster sample, we introduced a method for constructing full-sky Compton-\(y\) maps for EDE cosmologies, taking angular cluster correlation into account. To do so we utilise a line-of-sight integrated approach, using Limber's approximation, to convert the three dimensional power spectrum to the angular one, such that we neglect the correlation in redshift space. A Gaussian realisation of the angular power spectrum, normalised to the total number of expected clusters, leads to the expectation value of the number of clusters in a given pixel. We obtain then the spatial distribution of the cluster sample by drawing the number of clusters in a given pixel from a Poisson distribution, such that the mean of the distribution equals the cluster expectation value. Based on the full-sky SZ maps, we simulate \planck observations in all nine frequency channels, including the optical setup, detector noise, frequency response, as well as Galactic foregrounds. The resulting detected cluster samples have been compared to the fiducial \(\Lambda\)CDM case after cross-checking with the respective input catalogues, leading to the following results:
\begin{enumerate}
      \item We showed that the increased presence of galaxy clusters in EDE cosmologies is also reflected in the detected cluster sample, using the MFMF filtering technique. 
      \item The possibility of detection depends on the strength of the EDE. For the realistic CTNSDF case of EDE 3 there is no detectable effect on the redshift distribution relative to \(\Lambda\)CDM. It seems that a contribution of \(\bar{\Omega}_{\mathrm{DE,sf}}\sim 0.04\) is required for significant detection with the considered filtering method.  
      \item The presence of EDE also influences the contamination of the detected cluster sample via the relative noise contribution. For the EDE case the contamination is always lower than for the \(\Lambda\)CDM case. The effect of the relative noise level on the contamination is briefly discussed in the Appendix. 
\end{enumerate}
Under the assumption that \planck will be able to deliver all necessary information on foregrounds, it should be possible to detect significant deviations from the \(\Lambda\)CDM expectations, provided that the selection functions for the applied filtering methods are sufficiently known. And since our filtering method is optimised globally for full-sky application, it can be expected that methods working on patches outside the Galactic plane will find even more significant differences than our analysis. 

\begin{acknowledgements}
We are grateful to B. M. Sch\"afer for sharing with us the code for the filter construction and to Georg Robbers who
provided \(w(z)\) and the CMB power spectra for the EDE cosmologies. Special thanks go to Martin Reinecke for the marvellous C++ version of the \healpix package. This work was supported by the German \emph{Deut\-sche For\-schungs\-ge\-mein\-schaft, DFG\/} project Transregio TR33 The Dark Universe and by the \emph{IMPRS for Astronomy \& Cosmic Physics} at the University of Heidelberg.
\end{acknowledgements}

\bibliography{9990ref}

\appendix
\section{Impact of noise levels to the filter performance}
\begin{figure}
   	\centering
   	\includegraphics[width=0.9\linewidth]{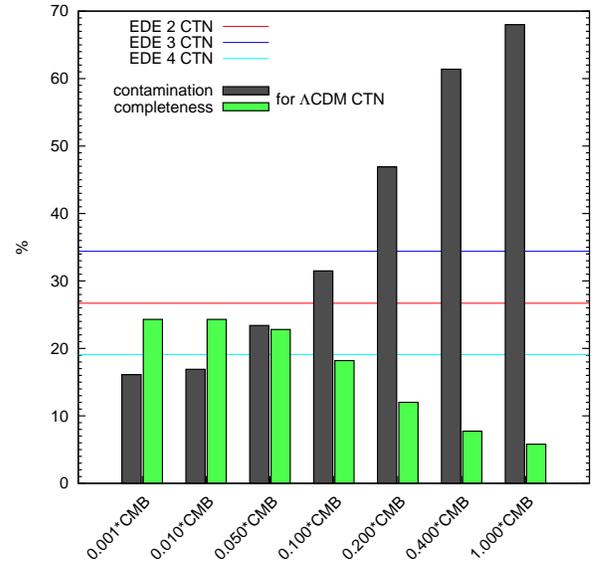}
   	\caption{Contamination (grey bars) and completeness (green bars) for the \(\Lambda\)CDM CTN dataset as function of the scaled CMB for a \(\mathit{Y}_{\mathrm{threshold}}=3\times10^{-4}\;\mbox{arcmin}^2\) and \(r_{\mathrm{search}}=15\;\mbox{arcmin}\). The solid lines show the according contamination levels for the CTN case of the models EDE 2, 3 and 4.}
        \label{Fig:contamin_study}
   \end{figure}

In order to study the surprising results in terms of the contamination of the detected cluster sample, we decided to perform a toy model test by downscaling the level of the CMB, which acts as source of noise to the filtering process. We applied in this case more strict association criteria of \(\mathit{Y}_{\mathrm{threshold}}=3\times10^{-4} \;\mbox{arcmin}^2\) and \(r_{\mathrm{search}}=15\;\mbox{arcmin}\) and included also the completeness, defined as the ratio of \(N_{\mbox{true}}\) over the number of cluster in the map above \(\mathit{Y}_{\mathrm{threshold}}\), in our analysis (denoted by the green bars in Fig. \ref{Fig:contamin_study}. By downscaling the CMB, one can mimic the effect of the relative noise level on the properties of detected cluster sample. The results of this test are shown in Fig. \ref{Fig:contamin_study}, where it can be seen that the contamination is decreasing with the decreasing noise level. The corresponding contaminations for the EDE 2, 3 and 4 models are illustrated by the solid lines and it can be seen that the levels of the EDE cosmologies can be reached by a relative lowering of the noise level. Hence, the presence of EDE does not only decrease the number of detectable clusters, but is also lowers the contamination by increasing the relative noise ratio between the background and the enhanced low and intermediate redshift cluster sample.  
%
%
\end{document}